\documentclass[article]{elsarticle}
\usepackage[utf8]{inputenc}  
\usepackage{multicol}
\usepackage{lineno}
\usepackage{hyperref}
\hypersetup{
	colorlinks,
	citecolor=blue,
	filecolor=blue,
	linkcolor=blue,
	urlcolor=blue,
}
\usepackage{subfig}
\usepackage{textcomp} 
\usepackage{xcolor}
\usepackage{ulem}
\usepackage{gensymb}
\usepackage{amsmath}
\usepackage{amsfonts}
\usepackage{amssymb}
\usepackage{epsfig}

\usepackage{graphicx}

\biboptions{sort&compress}

\journal{Journal of Nucl. Instrum. Methods Phys. Res. A}

\begin{document}
	\begin{frontmatter}
		
		\title{ Development and property study of the 12-\texorpdfstring{$\mu$}-m thick straw tubes with a diameter of 5 mm for the COMET Straw Tracker System}
        \author[GTU,JINR]{N.~Tsverava}
            \corref{corrauthor} 
            \cortext[corrauthor]{Corresponding author}
            \ead{Nikolozi.Tsverava@cern.ch}
		\author[GTU,SOKENDAI]{G.~Adamov}
		\author[GTU,JINR]{D.~Chokheli}
        \author[KEK]{H.~Nishiguchi}
		\author[GTU]{T.~Toriashvili}
		\author[GTU,JINR]{Z.~Tsamalaidze}

		\address[GTU]{Georgian Technical University, Tbilisi, Georgia}
		\address[JINR]{Joint Institute for Nuclear Research, Dubna, Russia}
		\address[KEK]{High Energy Accelerator Research Organization (KEK), Tsukuba, Japan}
		\address[SOKENDAI]{Graduate University for Advanced Studies (SOKENDAI), Tsukuba, Japan}
		
        \begin{abstract}
            The COMET experiment focuses on searching for the direct conversion of a muon into an electron on an aluminum nucleus without emitting a neutrino (so-called $\mu\rightarrow e$ conversion). This 
            conversion violates the lepton flavor conservation law, a fundamental principle in the Standard Model. The COMET experiment aims to achieve the muon-to-electron conversion sensitivity on a level of $10^{-17}$. The Straw Tracker System (STS) based on straw tubes could provide the necessary spatial resolution of 150 $\mu$m to achieve an momentum resolution for 105 MeV/c electrons better than 200 keV/c.

            The COMET experiment will be separated into two phases. Phase-I will operate with the 3.2-kW 8-GeV-proton beam, and Phase-II will operate with the beam intensity increased to 56 kW. The STS must operate in a vacuum with the inner pressure of 1 bar applied to straws. The initial design of 10-mm-diameter straws developed for Phase-I will not be as efficient with the 20 times higher beam intensity of Phase II, but the new STS design based on 5-mm-diameter 12-$\mu$m-thick straws could fully satisfy the required efficiency. The mechanical properties of these straws, such as sagging, displacement, and dependence of the diameter on overpressure, are discussed in this article.
          
        \end{abstract}

		
	\end{frontmatter}

    \pagebreak

\section{Introduction}
The COMET (COherent Muon to Electron Transition) experiment {\cite{comet_tdr2018}} is aimed to search for neutrinoless muon-to-electron conversion. This process lies beyond the Standard Model (SM) {\cite{pontecorvo47, SU5mu2e, kuno01, muhara13}}. The Straw Tracker System (STS) is an essential part of the COMET experiment. The main advantage of straw chambers is reduced multiple scattering due to an ultralight material with a low density in the charged particle propagation path, which increases the energy resolution for the particles, holding the momentum resolution better than 2\% for the 100-MeV region. This allows us to achieve a single-event sensitivity below $10^{-16}$ {\cite{hagime2017}}.

Historically, the straw tube system was first proposed as a tracking system for the Superconducting Super Collider (SSC, {\cite{straw4ssc, ssc_std}}) in 1989. And the trackers based on it were introduced for the COMPASS experiment {\cite{compass_lat}} and were widely used in experiments such as PANDA {\cite{panda_stt}}, ATLAS{\cite{atlas_trt}}, etc. The doubly wound helix (S-type) straws were used for these experiments. This type of straw is used to build the Straw-tube Tracker for the Mu2e experiment {\cite{mu2e_stt}.

Another type of straw (C-type) was introduced for the NA62 Low Mass Straw Tracker {\cite{na62_stt}} where the 10-mm-diameter tube shape was formed from a Mylar film with a thickness of 36 $\mu$m, and a seam along the straw tube was welded with an ultrasonic welder. This type of straw is used to build the Straw-tube System for COMET experiment {\cite{comet_sts}. The straws for COMET Phase-I are 10 mm in diameter, and the Mylar thickness is 20 $\mu$m.

In order to have a better resolution, smaller straws are very helpful from the material-budget point of view. Reducing the diameter and the thickness is the best solution to obtain better resolutions. The new design of the STS was proposed {\cite{hagime22}} based on 5-mm-diameter straw tubes for COMET Phase-II {\cite{Fukao:2021ior}}. These straws will be created using the 12 $\mu$m one-side aluminized Mylar film, thus reducing the overall thickness of the straw system.

\section{Acronyms, simplifications}
\begin{tabular} {p{0.35\linewidth} | p{0.6\linewidth}}
$\phi 5mm \times t 12 \mu m$ straw tube & 5-mm-diameter 12 $\mu $m-thick straw tube \\ 
\hline
overpressure & The relative pressure above atmospheric pressure delivered into the straw tube \\  
\hline
elongation & The axial displacement   
\end{tabular}

\section{The \texorpdfstring{$\phi 5mm \times t 12 \mu$}-m straw tube design and production methods}

Two different methods can be used to make thin-wall straw tubes. 
To produce S-type straws, two thin Mylar films with a selected metal coating (for instance, Al coating) are wound around a guiding axis as a helix, and glue was used to fix them (Fig. {\ref{fig:helixstrawtube}}}). The other method, C-type, is to make the straw tube from the Mylar tape by overlapping the edges of the tape and using the ultrasonic welder to stick these edges to each other (“direct adhesion straw” without glue) ({\ref{fig:ctypestrawtube}}).

A new method of ultrasonic welding technologies was developed at CERN and adapted by JINR for producing vacuum-compatible C-type straws {\cite{movchan2009, jinr_welding2016, nika_welding2019}}. One of the major advantages of the straw is its low gas leakage rate, which is essential for operation in a vacuum. For instance, the $CO_{2}$ gas leakage rate for long C-type straws welded using a 36-$\mu$m-thick Mylar film was observed to be $0.5\times10^{-3}$ ccm per m at 1-bar overpressure {\cite{movchan2009, jinr_welding2016}}. In comparison, the leakage rate for the S-type 15-µm-thick metalized Mylar straw with a diameter of 5 mm was less than $3.5\times10^{-3}$ ccm. In this case, the straw was filled with $CO_{2}$ gas at the pressure of 2 atm and inserted into the vessel filled with pure nitrogen gas {\cite{mu2e_stt2016}}.

    \begin{figure}[!htbp]
    	\centering
    	\subfloat[]{\includegraphics[height=0.15\textheight, keepaspectratio]
    		{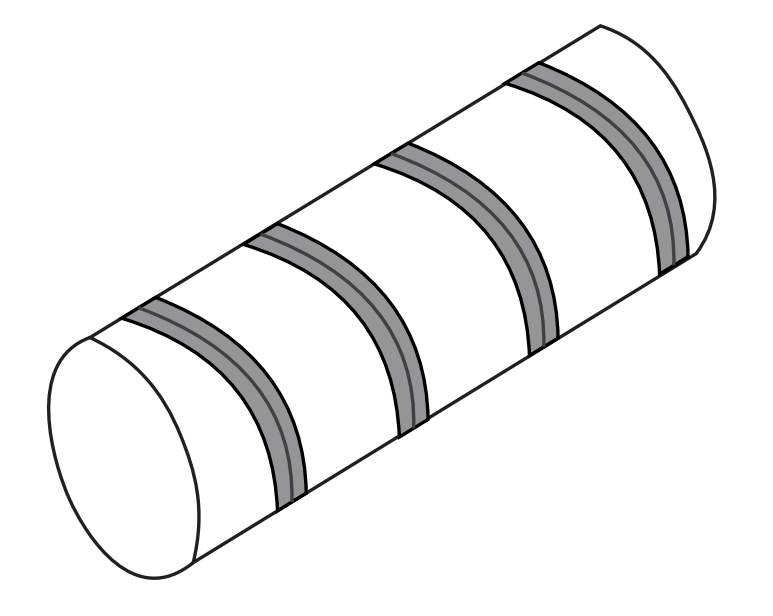} \label{fig:helixstrawtube}}
    	\subfloat[]{\includegraphics[height=0.15\textheight, keepaspectratio]
    		{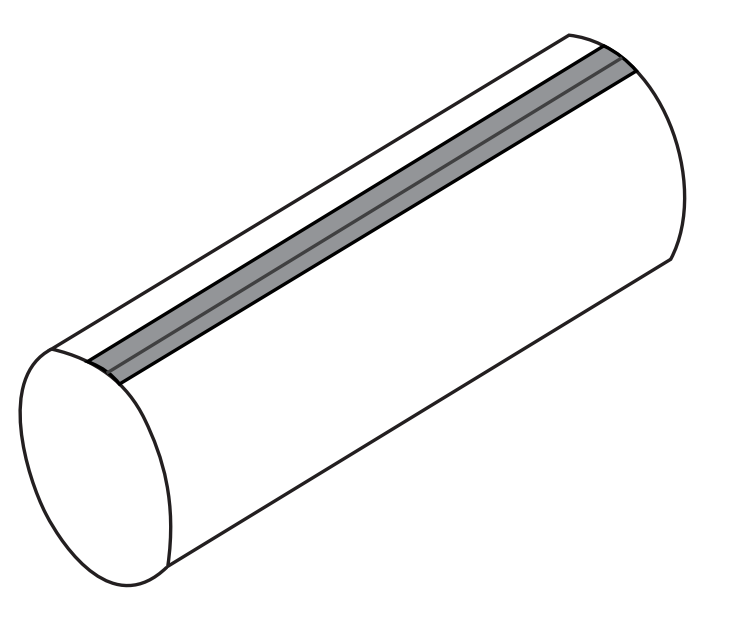} \label{fig:ctypestrawtube}}
    	\subfloat[]{\includegraphics[height=0.15\textheight, keepaspectratio]
    		{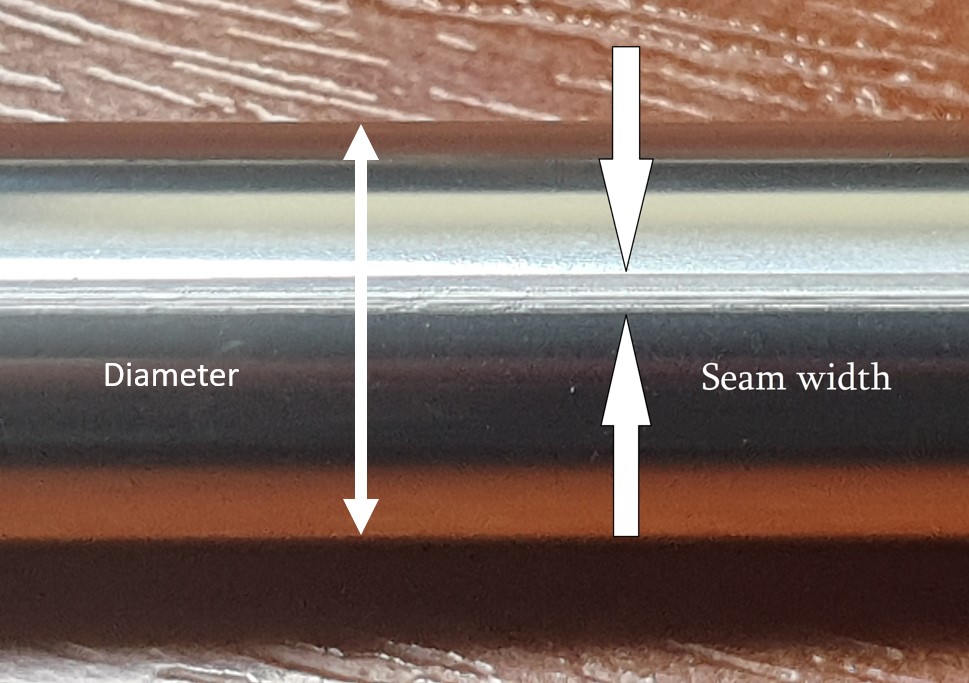} \label{fig:strawreal}}	
    	\caption{\small S-type straw tube (a); C-type straw tube with ultrasonic welding (b) and real C-type tube with the welding seam (c)}
    	\label{fig:strawtube}
    \end{figure}
This is the first time the 5-mm-diameter 12-$\mu m$ straw tubes have been produced using the ultrasonic welding technique. So, before creating the whole system with the proposed design, it was necessary to first create the straws according to the required parameters and test their mechanical properties.

\subsection{Desing of the \texorpdfstring{$\phi 5 mm \times t 12 \mu$}-m straw tubes}

The main component of the straw tracker for Phase II will be a $\phi 5 mm \times t 12 \mu$m Mylar straw tube {\cite{comet_tdr2018, hagime2017}} the inner side of which will be coated with a 70-nm layer of aluminum (cathode). According to the STS design, the straw length will be limited by 1.34 m. A gold-plated 20-$\mu$m-diameter tungsten wire (anode) will be positioned at its center, and high voltage will be applied to it. Straw tubes will be filled with the mixture of $Ar/C2H6(50:50)$ as a working gas under the excess pressure of 1 bar. All five modules of the tracker system will be located inside the superconducting magnets perpendicular to the magnetic field of 1 T and will operate in a vacuum below $10^{-6}$ mbar.

\subsection{Ultrasonic welding production station}

The C-type $\phi 5 mm \times t 12 \mu$m straw tube production station was built from scratch based on ultrasonic welding technology. An ultrasonic generator with the 44-kHz frequency (Fig. {\ref{fig:weldingstation}}) was used for the welding process.

    \begin{figure}[!htbp]
    	\centering
    	\subfloat[]{\includegraphics[height=0.15\textheight, keepaspectratio]
    		{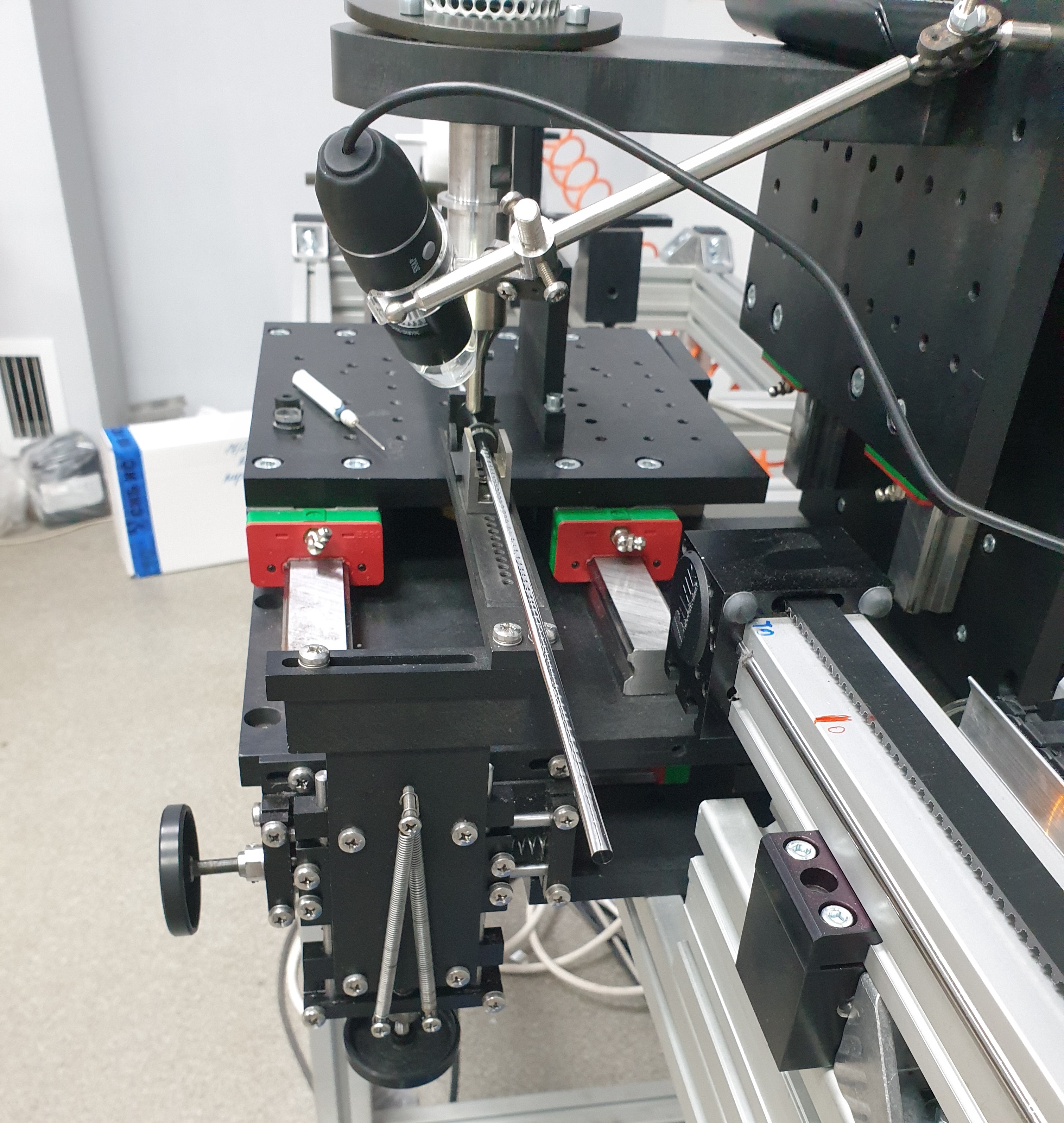} \label{fig: sourcenenergy}}
    	\subfloat[]{\includegraphics[height=0.15\textheight, keepaspectratio]
    		{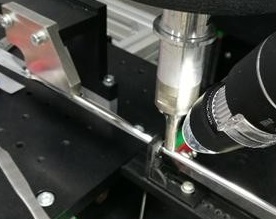} \label{fig: sourceDwattage}}
    	\subfloat[]{\includegraphics[height=0.15\textheight, keepaspectratio]
    		{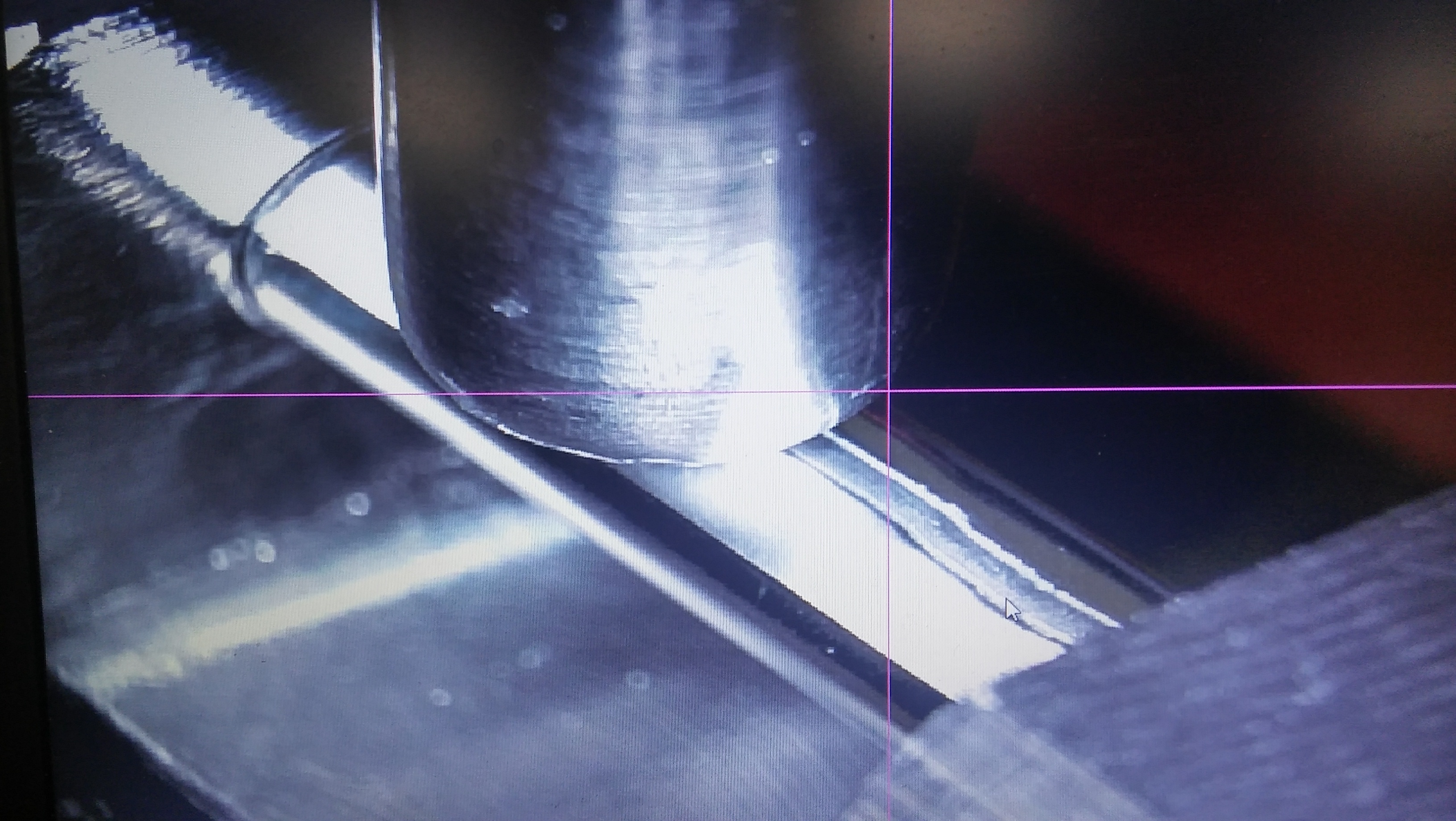} \label{fig: sourceDintrel}}	
    	\caption{\small Straw production machine (a); straw welding spot creation (b) and weld seam forming process (c)}
    	\label{fig:weldingstation}
    \end{figure}

During this process, the Mylar tape with a one-sided aluminum coating (cathode) is formed into a cylinder shape, and the overlapping edges are attached by ultrasonic welding in a straight line with no glue used (so-called “straight adhesion”).

It is necessary to monitor the seam size during straw production. It is empirically found that the seam size should be between 250 and 400 $\mu$m for such straws. A smaller size will increase the gas leakage. In contrast, a larger seam will affect the straw tube shape (Fig. {\ref{fig:seamsize}}). The gas leakage through the seam is caused by the destruction of the metalized inner layer of the tubes during production. The typical seam width for a $\phi 5 mm \times t 12 \mu$m straw is about 350 $\mu$m (Fig. {\ref{fig:seamwidth}}) and the seam thickness is around 22 $\mu$m (Fig. {\ref{fig:seamsideA}}, {\ref{fig:seamsideB}}).

    \begin{figure}[!htbp]
    	\centering
    	\subfloat[]{\includegraphics[width=0.33\textwidth, keepaspectratio]
    		{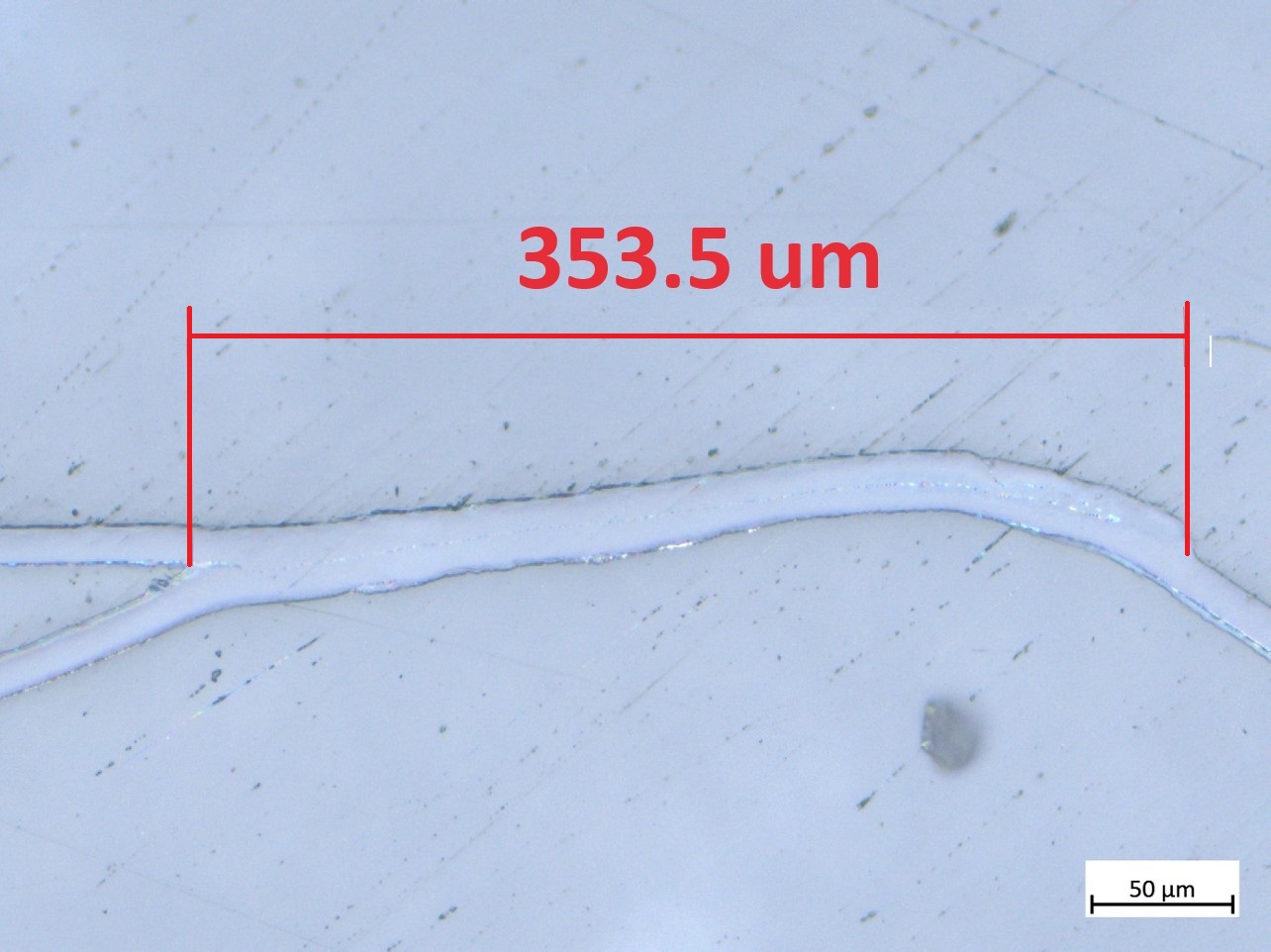} \label{fig:seamwidth}}
    	\subfloat[]{\includegraphics[width=0.33\textwidth, keepaspectratio]
    		{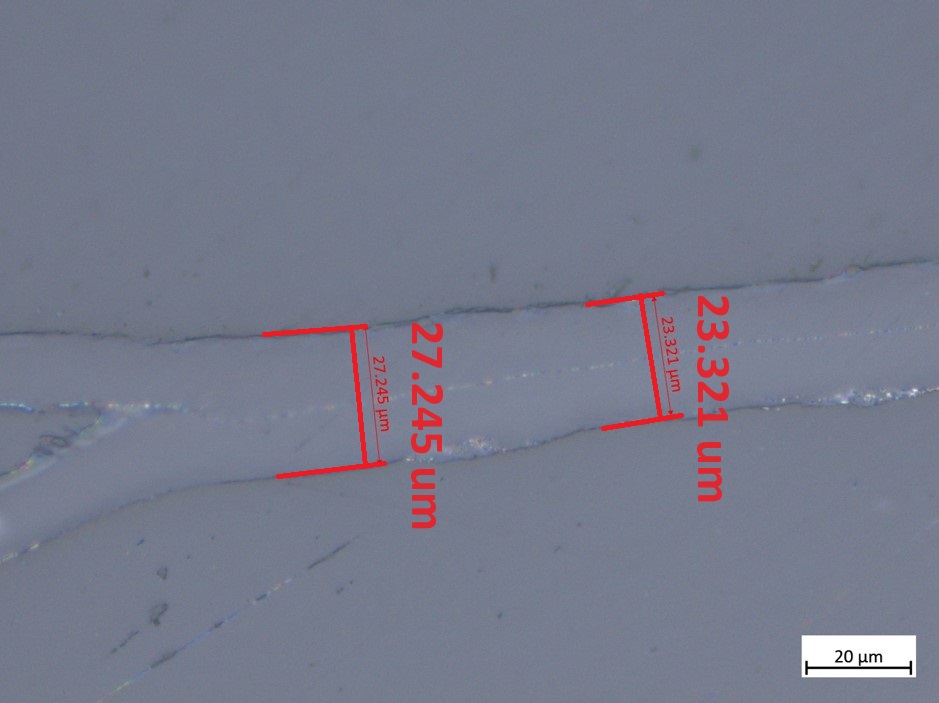} \label{fig:seamsideA}}
    	\subfloat[]{\includegraphics[width=0.33\textwidth, keepaspectratio]
    		{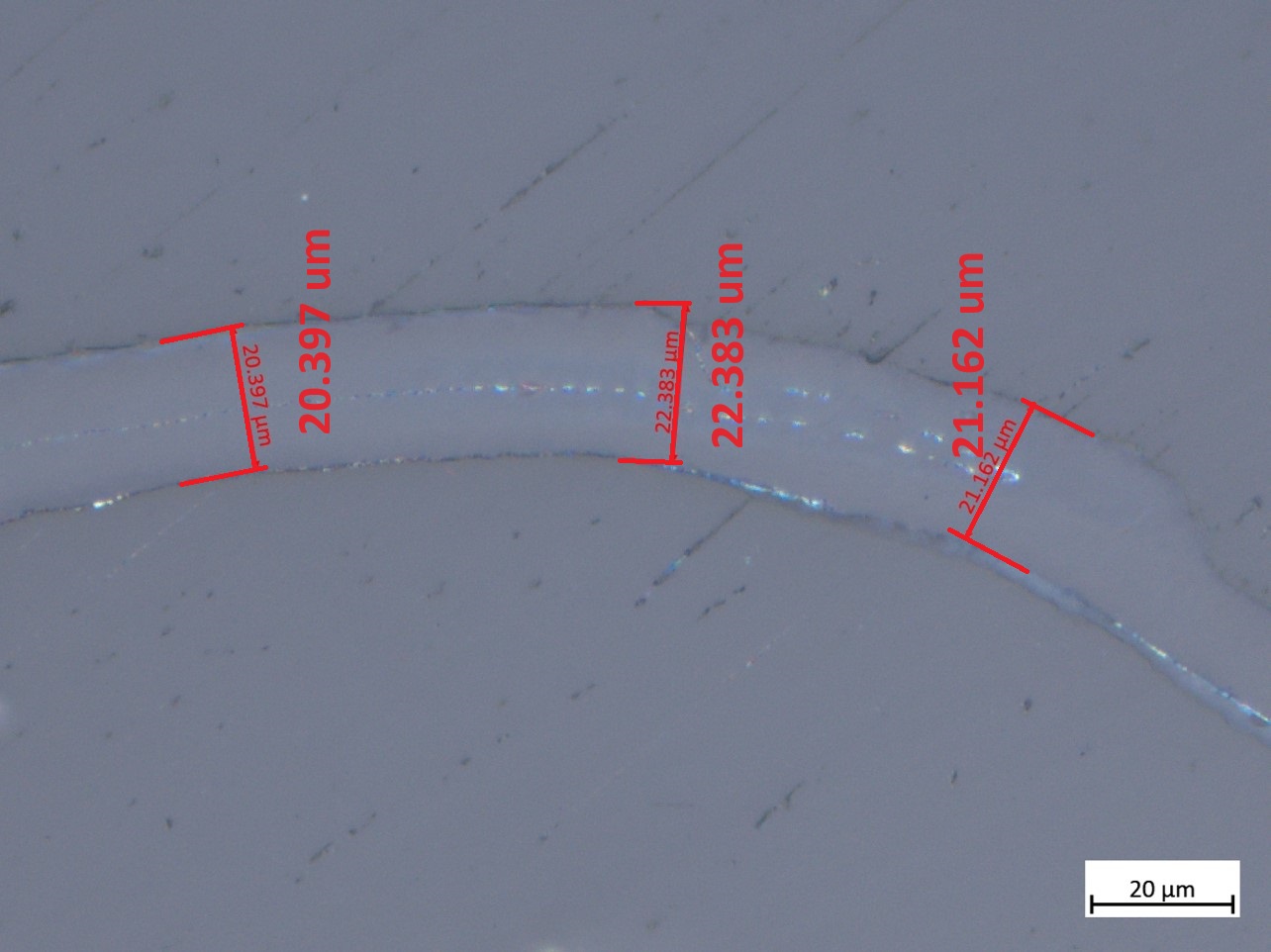} \label{fig:seamsideB}}	
    	\caption{\small The seam cross-section captured by microscope with 100x magnification (a); the left area of the seam (b); the right area of the seam (c) with 400x magnification}
    	\label{fig:seamsize}
    \end{figure}

A preliminary gas leakage test was performed on the 0.6-m-long straw. To perform it, a pure Ar gas with the 2-bar overpressure was delivered into the tube, and the pressure of the straw was continuously monitored for a week with a one-minute interval (Fig. {\ref{fig:leakweek}}). One end of the straw tube was directly attached to the wide range pressure gauge MPX5700 {\cite{mpx5700}} while the other end was hermetically sealed. Collected data was fitted using an exponential function. The slope of the fitting function indicates the pressure dropping rate, which was obtained to be 7.93 $\mu$Bar per minute. Using the Boyle–Mariotte law, we could calculate the gas leakage rate $\nu = dV/dt$ as \eqref{eq:leakage}:

\begin{equation} \label{eq:leakage}
	\nu = \frac{dV}{dt} =  \frac{ \frac{p_1}{p_{atm}} V_{str}  - \frac{p_2}{p_{atm}} V_{str}}{dt}  = \frac{V_{str}}{p_{atm}}\ \frac{dp}{dt} 
\end{equation}
Here, $p_{atm}$ is the absolute atmospheric pressure in bar, $V_{str}$ is the internal volume of the straw tube, and we do not take into account volume changes by pressure drop since it is negligible. The $dp/dt$ is the pressure-dropping rate.

Using equation \eqref{eq:leakage}, the distribution shown in Fig. {\ref{fig:leakweek}} was fitted by exponential function:
\begin{equation} \label{eq:leakagerate}
    p(t)
    = p_0 e^{-\nu \frac{p_{atm}}{V_{str}} t}
\end{equation}

The gas leakage rate was obtained to be $(0.92 \pm 0.01)\times10^{-4}$ ccm per minute for the 0.6-m-long straw. So, one can estimate the gas leakage for the $\phi 5 mm \times t 12 \mu$m straw tube at $1.55\times10^{-4}$ ccm per 1 m. The results of the careful and full-scale study of the gas leakage are the subject for another publication.

    \begin{figure}[!htbp]
    	\centering
    	\includegraphics[width=0.5\textwidth, keepaspectratio]{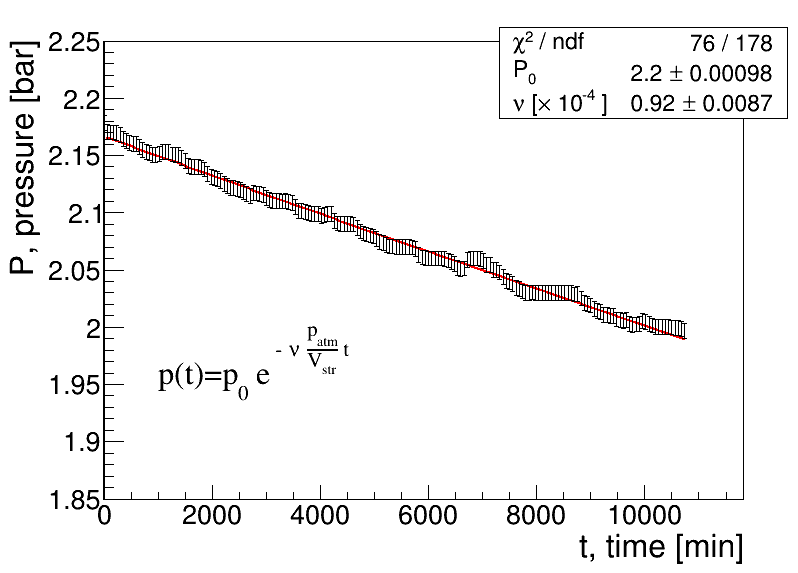}
    	\caption{\small Preliminary test of the gas leakage for the 0.6-m-long $\phi 5 mm \times t 12 \mu$m straw}
    	\label{fig:leakweek}
    \end{figure}

A 1.34-m-long $\phi 5 mm \times t 12 \mu$m straw was scanned for the diameter variations along the straw length to check the production quality. A Micro-Epsilon “optoCONTROL 2600” laser micrometer was used for this study {\cite{opto2600}}. It was moved along the straw with a constant speed and pictured the diameter with equal time intervals, thus measuring every 10 $\mu$m along the straw. Variations in the straw diameter for different overpressure in the longitudinal direction are presented in Fig. {\ref{fig:diam1}} (the uncertainties for these measurements are presented in Tab. \ref{tab:errors}, subsection \ref{errors}). A distribution of the diameter variation obtained at each pressure was approximated by the Gaussian fit function. It was found that the standard deviations for the diameter for the cases where overpressure delivered into the straw are $\sigma<4 \mu m$ (Fig. {\ref{fig:diam2}}). And FWTMs for these distributions did not exceed 18 $\mu$m. As expected, even the overpressure of 0.5 bar effectively helps to mold the cross-section shape of the straw close to the circle in contrast to the case with no overpressure.

    \begin{figure}[!htbp]
    	\centering
    	\subfloat[]{\includegraphics[width=0.5\textwidth, keepaspectratio]
    		{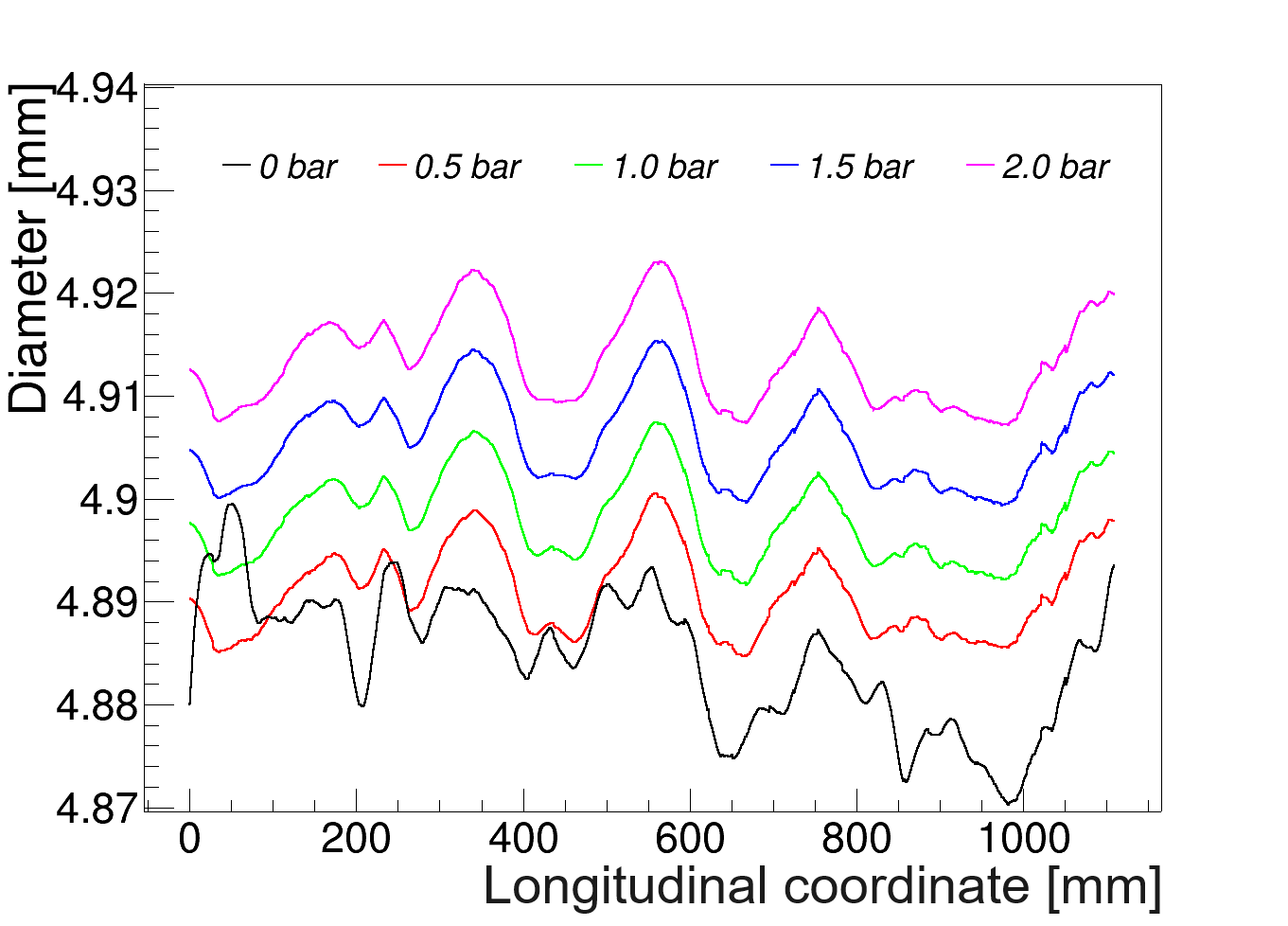} \label{fig:diam1}}
    	\subfloat[]{\includegraphics[width=0.5\textwidth, keepaspectratio]
    		{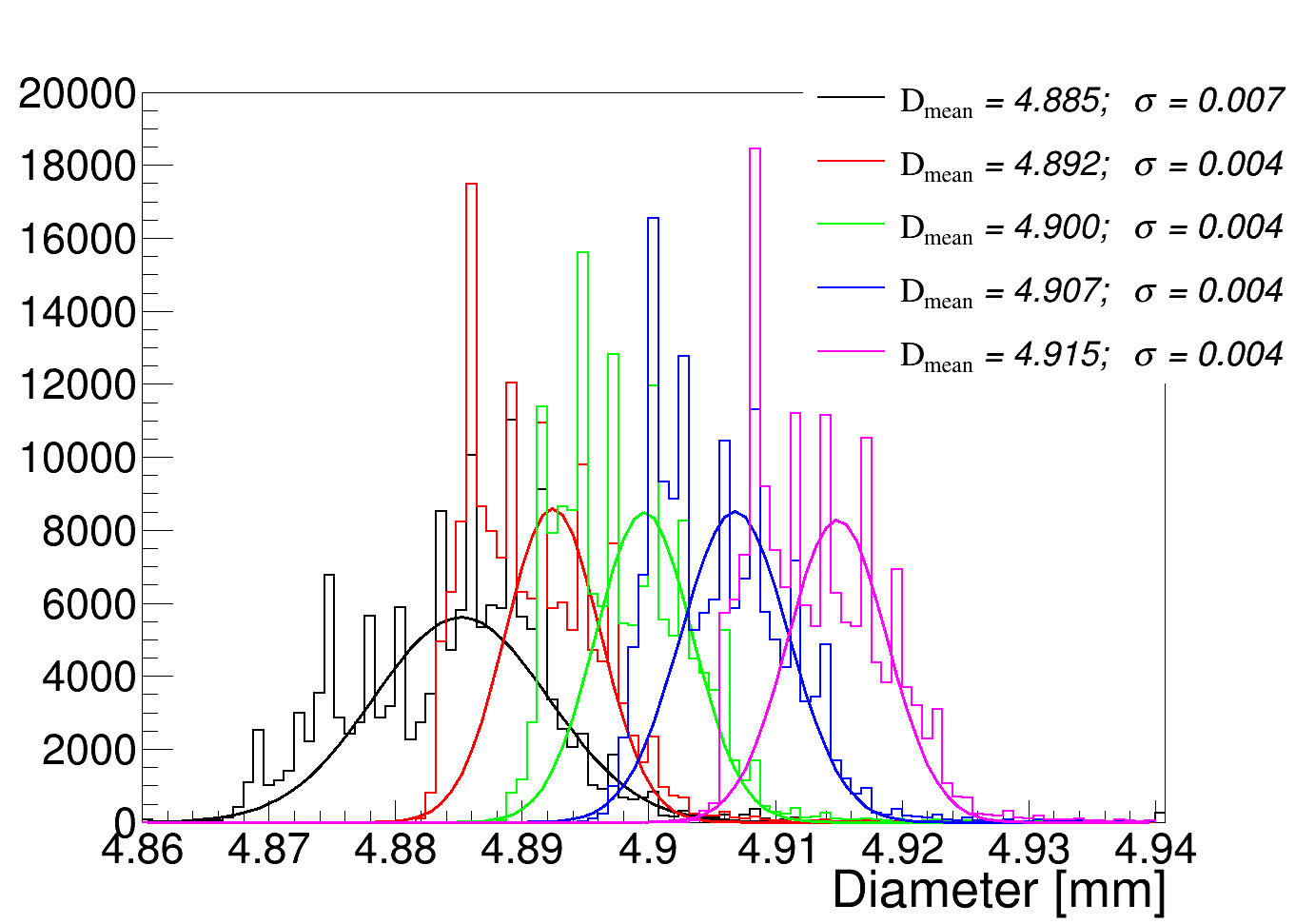} \label{fig:diam2}}
    	\caption{\small Diameter variations along the straw for different overpressure. The scanning area for the 1.34-m-long straw was 1100 mm with a step of 10 $\mu$m.}
    	\label{fig:longdiavar}
    \end{figure}

\section{  \texorpdfstring{$\phi 5 mm \times t 12 \mu$}-m straw tubes mechanical properties study}

About 100 1.34-m-long $\phi 5 mm \times t 12 \mu$m straw tubes were fabricated to check production stability. 

The mean value of the diameter for this bunch of straws was measured at the ends of a straw once it was pressurized by 3 bar and was found to be $\phi4.94\pm0.02\ $mm. As a next step, the mechanical properties of this type of straw should be studied. We provided the complete set for a study of the mechanical properties, such as straw tube elongation as a function of the load, straw tube elongation caused by the overpressure, straw tube diameter dependence on the load and on overpressure, bending of the pressured straw tube in the vertical position, and sagging of the horizontally placed straw caused by gravitation. For instance, precise measurements for 9.8-mm-diameter straws with a wall thickness of 36 $\mu$m and 18-mm-diameter straws with a wall thickness of 54 $\mu$m (both C-type straws) were done in {\cite{glonti2018}}.

\subsection{Measurements setup}

Two unique setups were built to study mechanical properties using standard T-slot profiles of 45 series. Some of the measurements were performed on setup $\#1$, which was equipped with a 1-X translation stage holding tension gauge on the one end of the straw tube while the other end was just fixed (Fig. {\ref{fig:setup1}}). The rest of the measurements were done on setup $\#2$, which was created from scratch to provide wider tests of the straw tubes (Fig. {\ref{fig:setup2}}). This new setup allowed us to study the mechanical properties of the tubes with lengths up to 4.5 m, which could be set at different angles from 0 to 90 degrees, etc.

    \begin{figure}[!htbp]
    	\centering
    	\subfloat[]{\includegraphics[height=0.15\textheight, keepaspectratio]
    		{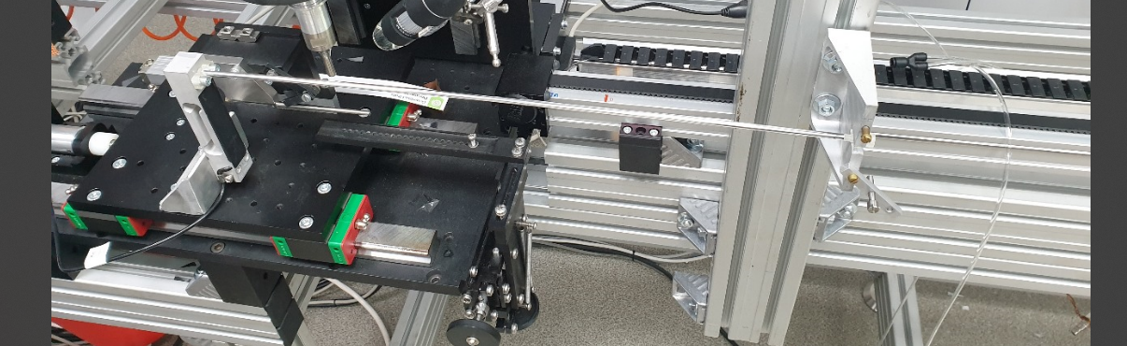} \label{fig:setup1}}
    	\subfloat[]{\includegraphics[height=0.15\textheight, keepaspectratio]
    		{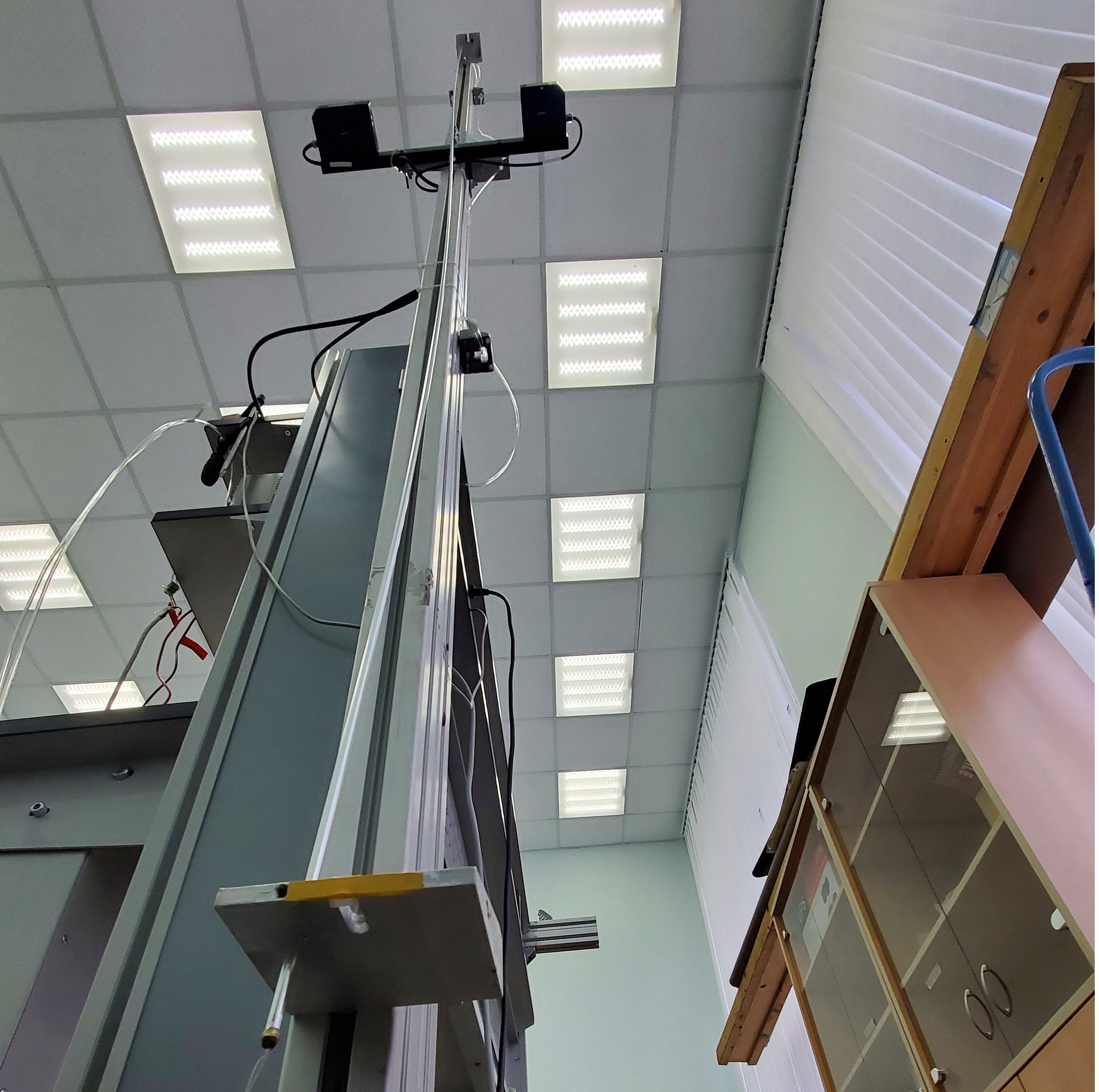} \label{fig:setup2}}
    	\caption{\small Special setups built to study the mechanical properties: setup $\#1$ (a) and setup $\#2$ (b)}
    	\label{fig:setup}
    \end{figure}

To measure the straw diameter variation, sagging and bending, a MicroEpsilon “optoCONTROL 2600” laser micrometer was used {\cite{opto2600}}. The sensor measurement range was based on the width of the laser beam and corresponded to 40 mm with an accuracy of 1 $\mu$m (Fig. {\ref{fig:opto2600}}). This laser micrometer could operate in several modes, and we only used two of them: “Mode 1” to measure the position of the light/dark edge in absolute values and “Mode 2” to calculate the difference between two light/dark edges in relative values. The first mode allowed measuring the straw tube elongation from the initial position, and the second mode allowed monitoring the straw tube diameter during the measurements. This device was installed in the middle of the straw being measured.

    \begin{figure}[!htbp]
    	\centering
    	\includegraphics[height=0.20\textheight, keepaspectratio]{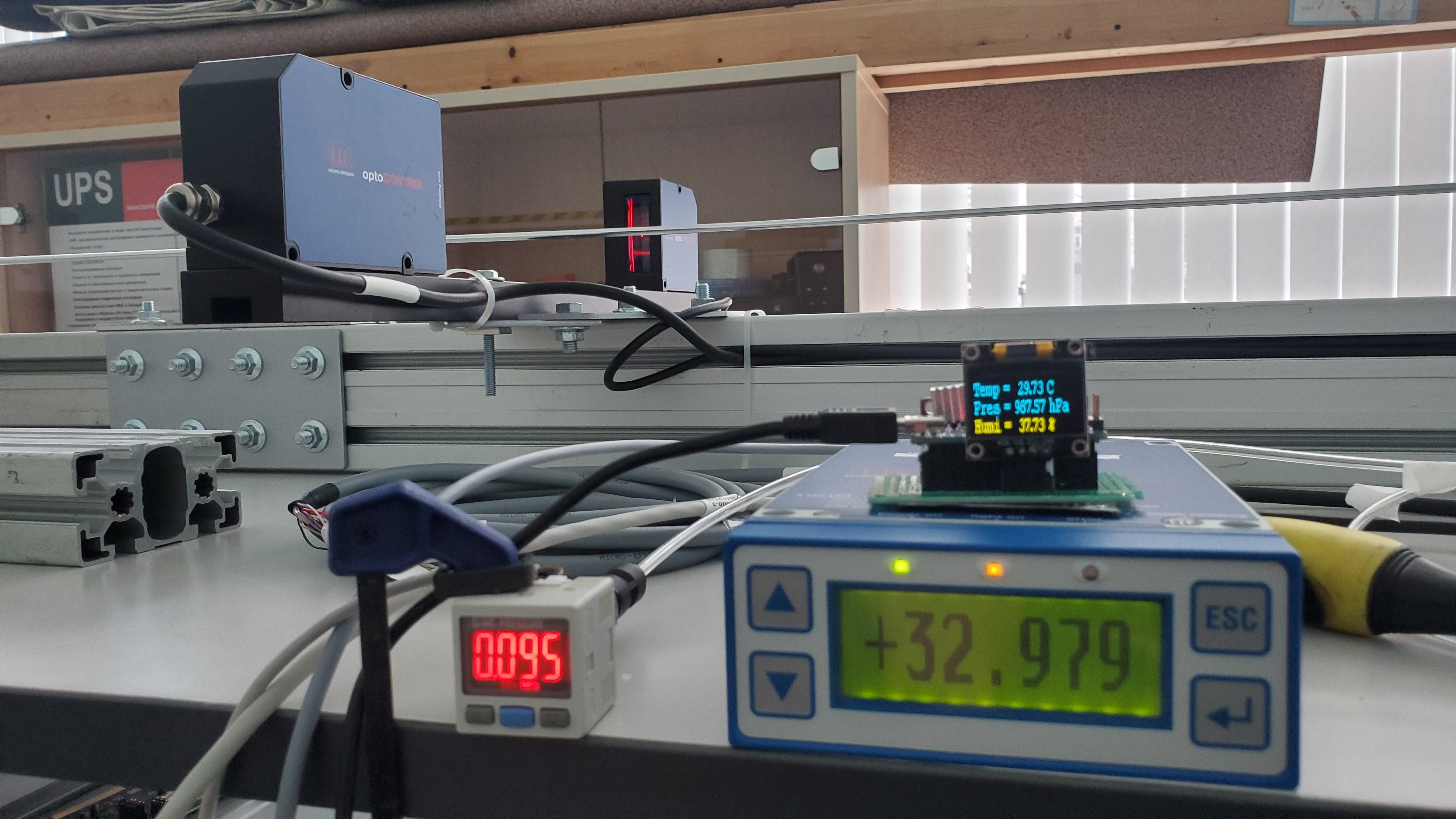}
    	\caption{\small The Micro-Epsilon “optoCONTROL 2600” laser micrometer measures a straw elongation from the initial position in absolute values}
    	\label{fig:opto2600}
    \end{figure}

The elongation measurements on setup $\#$2 were done using the following technique. The narrow laser beam was spread at a wide angle using the de-focusing lens. A small paper strip marker was placed on the end of the straw. The straw end with the marker was placed in the middle position from the laser source to the screen with a ruler. The light beam was shadowed by a marker, and we could monitor the variation of the straw length on the screen using a ruler (Fig. {\ref{fig:laserelong}}).

    \begin{figure}[!htbp]
    	\centering
    	\subfloat[]{\includegraphics[height=0.15\textheight, keepaspectratio]
    		{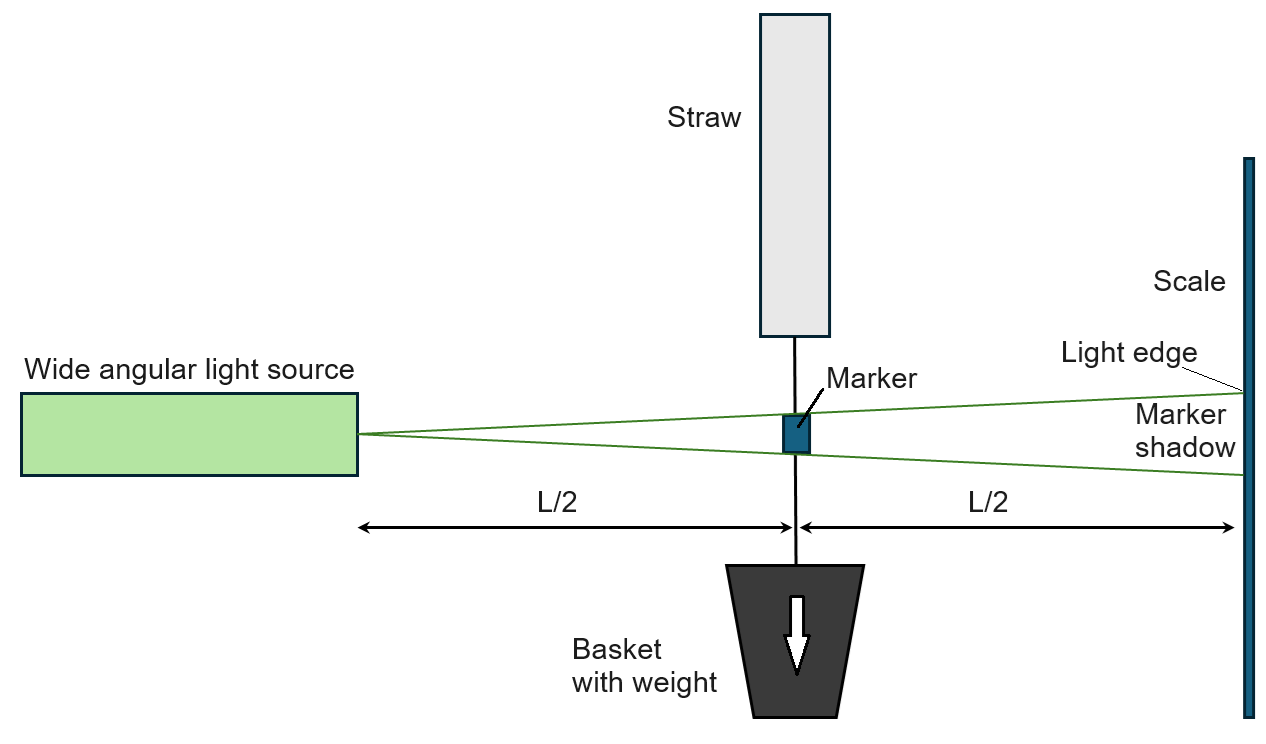} \label{fig:laserelong1}}
    	\subfloat[]{\includegraphics[height=0.15\textheight, keepaspectratio]
    		{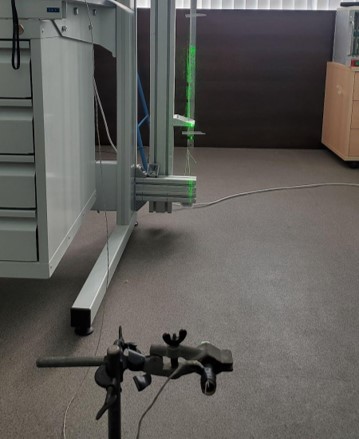} \label{fig:laserelong2}}
    	\caption{\small Sketch of the elongation measurements using setup $\#$2 (a) and a view of the setup (b)}
    	\label{fig:laserelong}
    \end{figure}

\subsection{Measurements uncertainties} \label{errors}

Initial accuracy and uncertainties used to determine the measurement precisions are presented in Tab. \ref{tab:errors}:
\begin{table}[ht]
\caption{Initial accuracy and uncertainties}
\begin{tabular} { p{0.85\linewidth} | p{0.13\linewidth}}
Accuracy in weight measured by tension gauge & 1 g\\ 
\hline
Accuracy in weight measured by scale & 4 g\\ 
\hline
Accuracy in length measured by 1-D translation stage & 0.1 mm\\  
\hline
Accuracy in length measured by scale ruler with precision marking & 0.25 mm\\  
\hline
Accuracy in pressure measured by pressure gauge & 0.05 bar\\  
\hline
Uncertainties in displacement based on laser micrometer 
(std. dev. for averaging of 1000 measurements)
 & 0.005 mm \\
\hline
Uncertainties in diameter based on laser micrometer 
(std. dev. for averaging of 1000 measurements)
 & 0.005 mm \\

\end{tabular}
 \label{tab:errors}
\end{table}

\subsection{Study of \texorpdfstring{$\phi 5 mm \times t 12 \mu$}-m straw tube elongation as a function of the external load}

The straw tube elongation tests were done for 0.5, 1.0, and 1.34-meter-long tubes under various loads.

The elongation test of 0.5- and 1.0-m-long tubes was done on setup $\#$1 (Fig. {\ref{fig:setup1}}) when one end of the straw tube was fixed on the tension gauge and the other end was just constantly fixed. The tension gauge was installed on the 1-X translation stage. The gauge was connected to the computer for remote data acquisition and control. Measurements with these tubes were done with no overpressure delivered to tubes.

The elongation within the elastic area could be calculated according to Hooke’s law \cite{sopromat}: 
\begin{equation} \label{eq:hooke}
	\delta L = \frac{F}{K_a} = \frac{mg}{K_a}
\end{equation}
here, $\delta L$ is the linear elongation of the tube, $F=mg$ is the axial load, and $K_a$ is the axial stiffness of the straw. And the relative (relative deformation) (strain) $\varepsilon$ could be defined as:
\begin{equation} \label{eq:strain}
	\varepsilon = \frac{\delta L}{L} = \frac{\sigma_t}{E_m}
\end{equation}
here, $L$ is the length of the straw, $\sigma_t$ is the stress (force per unit area), and $E_m$ is Young’s modulus for the straw tube. 

At the same time, the axial stiffness is inversely proportional to the straw tube length and could be calculated as:
\begin{equation} \label{eq:stiff}
	K_a = \frac{F}{\delta L} = \frac{\sigma_z S_{cr}}{\delta L} = \frac{\varepsilon E_m S_{cr}}{\delta L} = \frac{\delta L}{L}\frac{E_m S_{cr}}{\delta L} = \frac{E_m S_{cr}}{L} = \frac{\pi TD}{L} E_m
\end{equation}
here $S_{cr}= \pi(D^2-(D-2T)^2)/4 \approx \pi DT$ is the cross-section of the thin-walled straw tube perpendicular to the load, and $\sigma_z=\varepsilon E_m$ is the tensile/longitudinal stress (force per unit area). Note that the variation of the straw cross-section produced by the load was negligible, and we did not consider it.

Combining these two equations (\eqref{eq:hooke} and \eqref{eq:stiff}), we could calculate the elongation of the straw end under the external axial load as:
\begin{equation} \label{eq:elong}
	\delta L = \frac{mg}{K_a} = \frac{gL}{\pi TD} \frac{m}{E_m}
\end{equation}

Figure {\ref{fig:elongVSload}} represents the results of these measurements for 0.5- and 1.0-m-long straw tubes for various overpressures: without excess pressure, and with overpressures of 0.5 and 1.0 bar for finding the difference in stiffness by the external load for different overpressures. The graphs were fitted with function \eqref{eq:elong} to prove that the elongation was within the elastic area according to Hooke’s law. 

    \begin{figure}[!htbp]
    	\centering
    	\subfloat[]{\includegraphics[width=0.49\textwidth, keepaspectratio]
    		{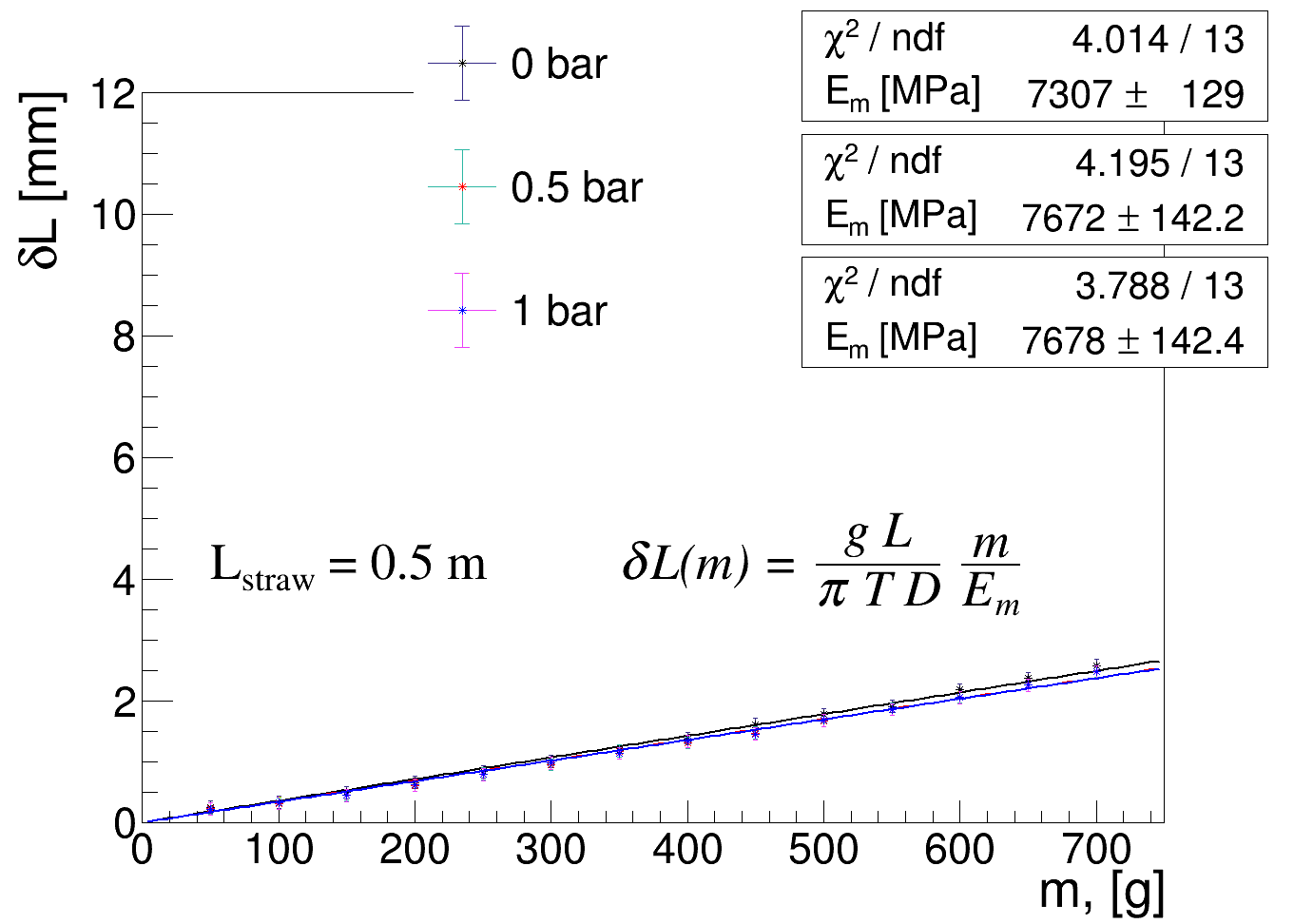} \label{fig:elongVSload1}}
    	\subfloat[]{\includegraphics[width=0.49\textwidth, keepaspectratio]
    		{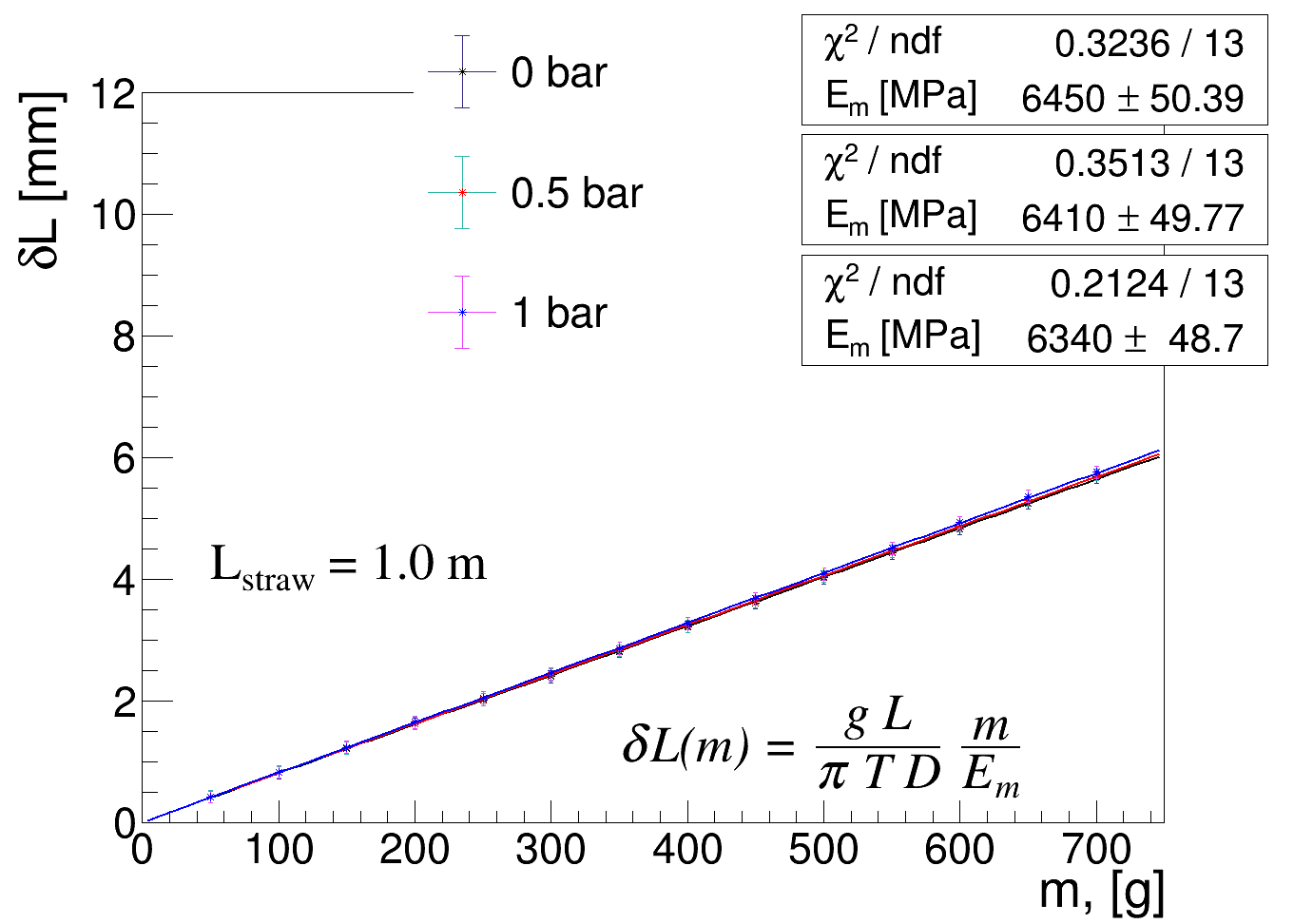} \label{fig:elongVSload2}}
    	\caption{\small Elongations for 0.5- (a) and 1-m-long (b) $\phi 5 mm \times t 12 \mu$m straw tube}

    	\label{fig:elongVSload}
    \end{figure}

The elongation tests for the 1.34-m-long straw tube were done on setup \#2 (Fig. {\ref{fig:setup2}}). In this case, a physical weight was used to stress the tube, which was hung vertically. A marker was stuck on the end of the straw tube where the weight basket was installed. A laser widespread light source was used to illuminate the marker, and the position of the marker shadow was monitored on the screen with a scale on it located behind the straw tube (Fig. {\ref{fig:laserelong}}). The straw was placed halfway from the light source to the screen with the scale. The values of the positions were read out by a camera with 25x magnification installed in front of the screen.

These tests were carried out without excess pressure and at overpressures of 0.5 and 1.0 bar to find the difference in stiffness by the external load for different overpressures. The resulting graphs for the elastic area are shown in Fig.  {\ref{fig:elongVSloadVspress1}}. Again, according to Hooke’s law, both graphs were fitted with function \eqref{eq:elong} for the elastic area. The area, including non-elastic deformation of the straw tube for the case of 1 bar delivered, is shown in Fig. {\ref{fig:elongVSloadVspress2}}.

    \begin{figure}[!htbp]
    	\centering
    	\subfloat[]{\includegraphics[width= 0.5\textwidth, keepaspectratio]
    		{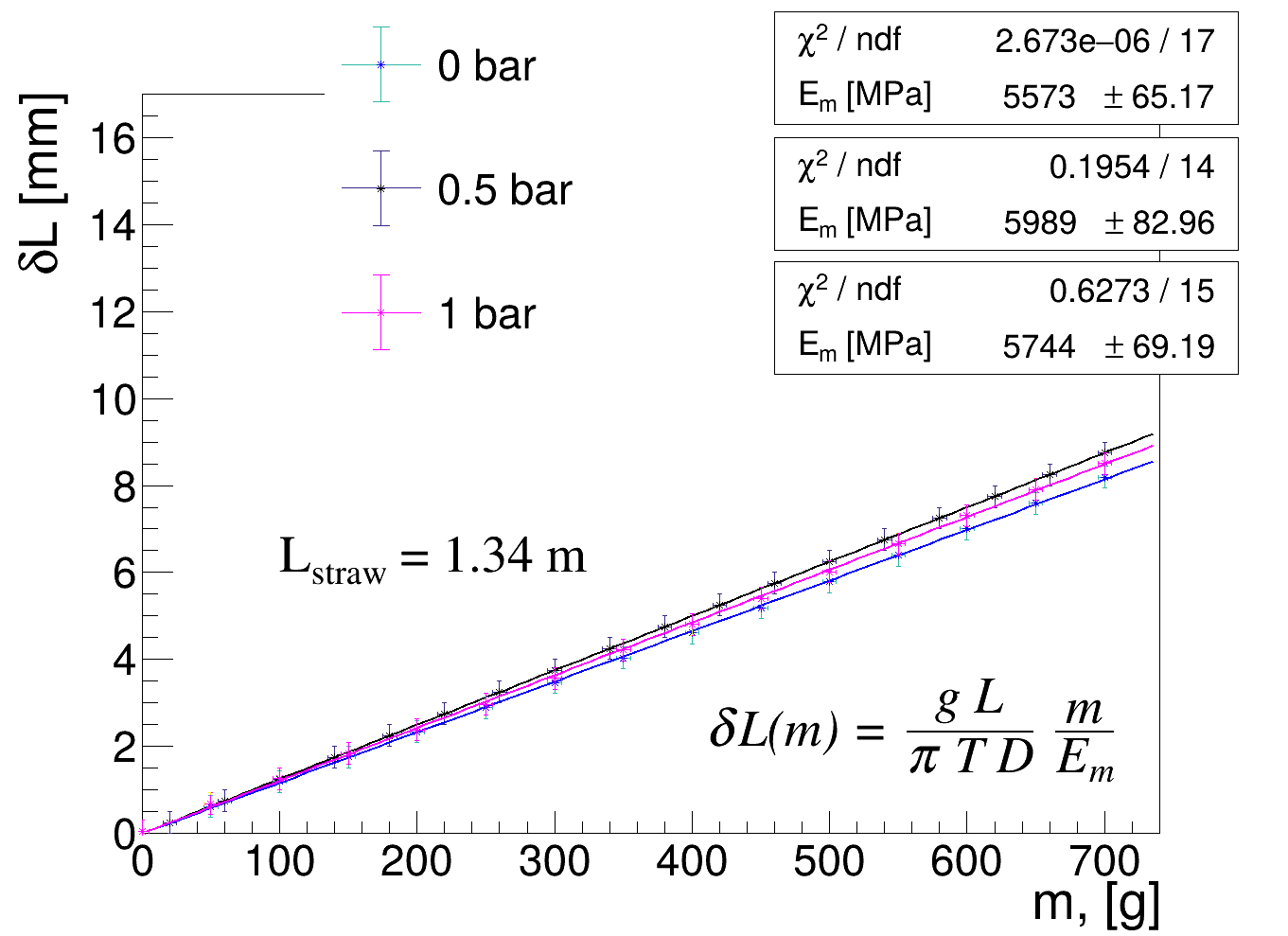} \label{fig:elongVSloadVspress1}}
    	\subfloat[]{\includegraphics[width=0.5\textwidth, keepaspectratio]
    		{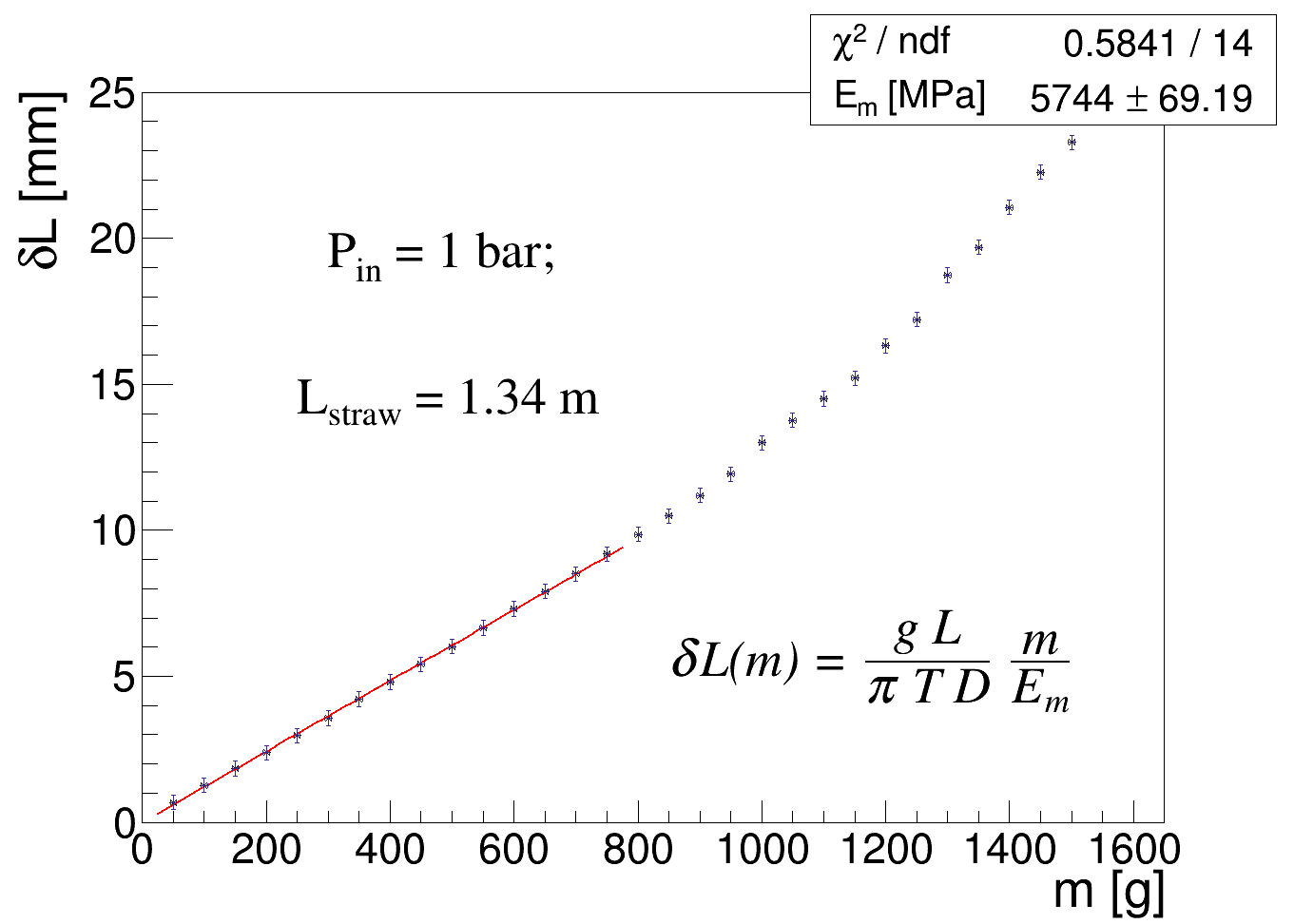} \label{fig:elongVSloadVspress2}}
    	\caption{\small Elongations for 1.34-m-long $\phi 5 mm \times t 12 \mu$m straw tube (a) for various overpressure inside of the elastic area and the area including non-elastic deformation of the straw tube at 1.0 bar overpressure delivered (b). 
        }
    	\label{fig:elongVSloadVSpress}
    \end{figure}

In the case of 1.0-bar overpressure, the elastic area was limited by the 700-g load (Fig. {\ref{fig:elongVSloadVspress2}}). Obviously, this limitation results from the combined stress of the external load and the internal pressure.

The combined value for Young’s modulus should be $E_m = 6.57 \pm 0.89$ GPa in comparison to 5.24 GPa ($0.76 \times 10^6$ Psi) obtained for Mylar tapes in {\cite{Becker1965ElasticMO}}.

\subsection{Straw tube elongation by overpressure}
In addition to the elongation caused by the external load attached to the tube ends, it is also important to know how the overpressure delivered into the straw acts on elongation. So, we measured elongation as a function of the straw tube overpressure in the 0.5- and 1.0-m-long tubes. This study was done on setup \#2. 

The free-sealed end of the straw begins to stretch along the axis of the tube while the other end is fixed. Special rings are installed every 10 cm along the length of the straw tube to prevent it from bending. An optoCONTROL 2600 laser sensor was used to measure elongation, which operated in “Mode 1” and monitored the free end of the straw.

Elongation $\delta L_p$  of the straw caused by overpressure consists of two parts: the elongation of the straw $\delta L_{ax}$ caused by overpressure acting on the plugs at the straw ends and the shortening of the straw $\delta L_{sh}$ according to Hooke's law when overpressure is acting on the straw wall. 

The elongation of the straw $\delta L_{ax}$ caused by overpressure acting on the plugs at the straw ends using equation \eqref{eq:strain} could be calculated as:
\begin{equation} \label{eq:elongpressplug}
    \delta L_{ax} = \varepsilon L = \frac{\sigma_z}{E_m}L = \frac{F_{p}}{S_{cr}}\frac{L}{E_m} =
    \frac{\pi D^2p}{4\pi TD} \frac{L}{E_m} = \frac{D L}{4T} \frac{p}{E_m}
\end{equation} 
here, $F_p=p S_p$ the force acting on the straw ends because of overpressure, $p$ is the overpressure, $S_p=(\pi (D-2T)^2)⁄4 \approx (\pi D^2)⁄4$ is the plug area, and $S_{cr}= \pi(D^2-(D-2T)^2)/4 \approx \pi DT$ is the cross-section of the thin-walled straw tube. 

To calculate the shortening of the straw $\delta L_{sh}$ when overpressure is acting on the straw wall, we need to calculate a tangential stress $\sigma_t$ of the straw:
\begin{equation} \label{eq:stress_t}
    \sigma_t = \frac{F_t}{S_t} = \frac{p S_{ax}}{S_t} = \frac{L(D-T)}{2LT} p= \frac{D-T}{2T}p \approx \frac{D}{2T}p
\end{equation} 
here $F_t=pS_{ax}$ is the tangential force on the axial section, $S_{ax}=L(D-T)$ is the axial area of the straw tube, and $S_t=2LT$ is the wall area on the axial section.

The axial deformation of the straw caused when overpressure is acting on the straw wall could be calculated as
\begin{equation} \label{eq:elong_ax}
    \delta L_{sh} = \frac{\sigma_z}{E_m}L = -\nu \frac{\sigma_t}{E_m}L = -\nu \frac{DL}{2T} \frac{p}{E_m}
\end{equation}
here, $\nu = - \sigma_z /\sigma_t$ is Poisson’s ratio for the straw tube. 

By combining equations \eqref{eq:elongpressplug} and \eqref{eq:elong_ax}, elongation of the straw by overpressure could be calculated as:
\begin{equation} \label{eq:elong_press}
    \delta L_{p} =\delta L_{ax} + \delta L_{sh}= \frac{D L}{4T}  - \nu \frac{D}{2T} \frac{p}{E_m} = \frac{D L}{4T}(1-2\nu) \frac{p}{E_m}
\end{equation}

The results of the elongation caused by overpressure are presented in Fig. \ref{fig:elongVSpress}. Both graphs were fitted with function \eqref{eq:elong_press} for the elastic area. Poisson's ratio for the Mylar tape was taken to be 0.4.

\begin{figure}[!htbp]
    \centering
    \subfloat[]{\includegraphics[width=0.5\textwidth, keepaspectratio]
        {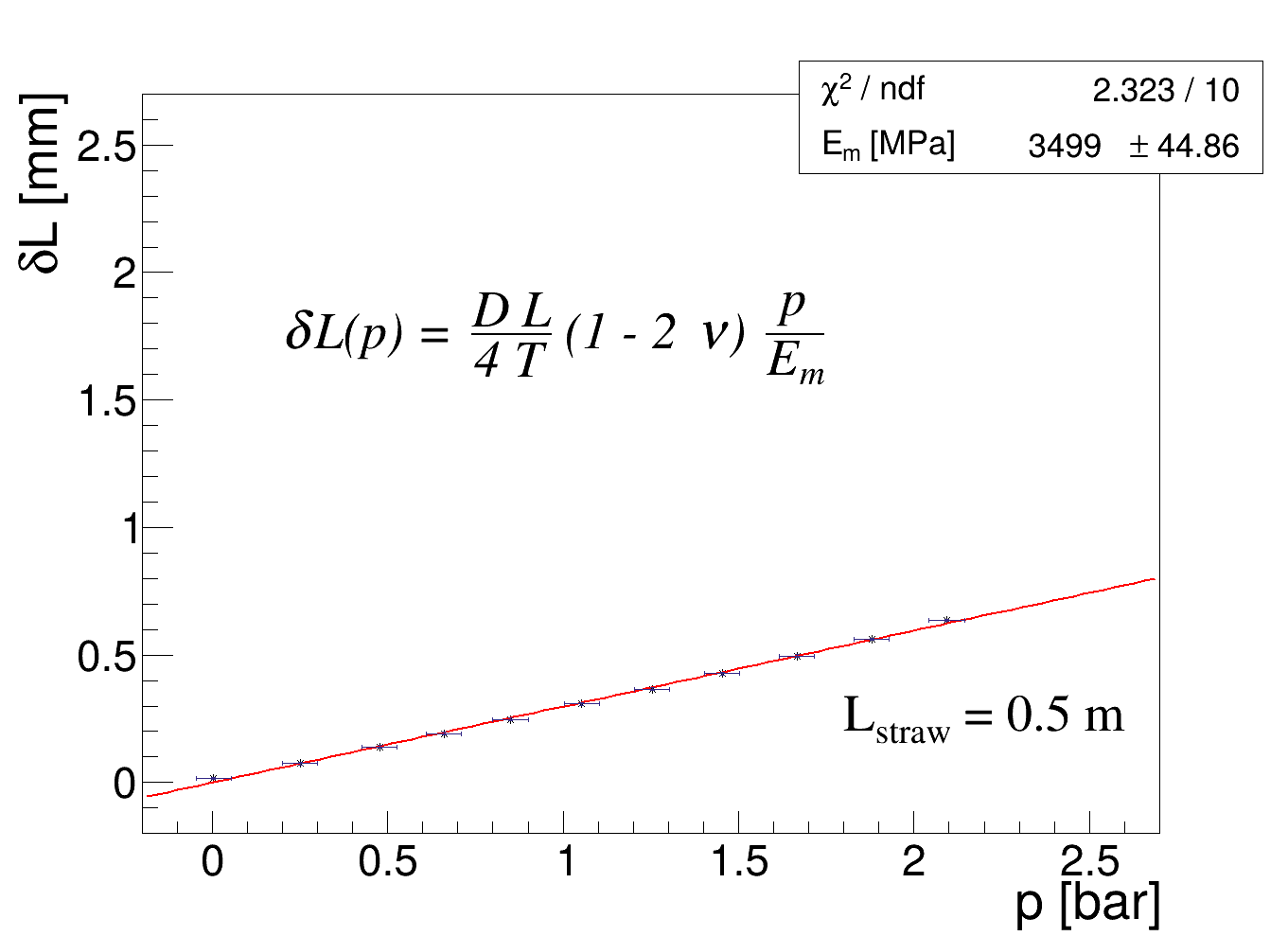} \label{fig:elongVSpress1}}
    \subfloat[]{\includegraphics[width=0.5\textwidth, keepaspectratio]
        {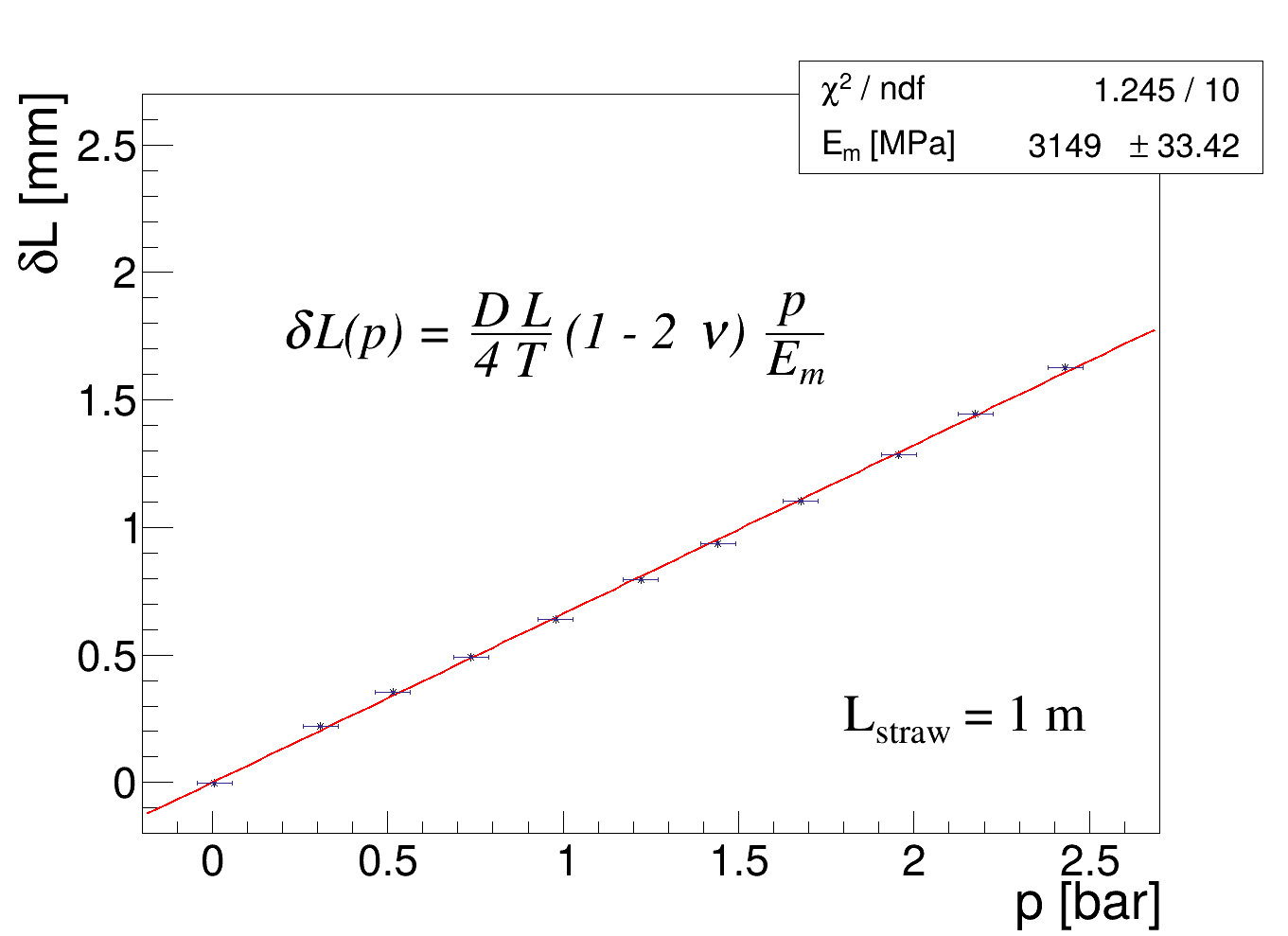} \label{fig:elongVSpress2}}
    \caption{\small Elongation for the 0.5-m-long (a) and 1.0-m-long (b) $\phi 5 mm \times t 12 \mu$m straw tube as a function of the overpressure. No load attached.}
    \label{fig:elongVSpress}
\end{figure}

It was found that the elongation of the 0.5-mm-long straw tube is $0.31 \pm 0.03$ mm per bar and $0.66 \pm 0.04$ mm per bar for the 1-m-long straw. Extrapolating this result to the 1.34-m-long straw tube, we could estimate the elongation for the 1.34-m-long straw tube caused by overpressure as 0.9 mm per bar.

\subsection{Diameter as a function of load and overpressure}

Another study was to measure the straw tube diameter $D(F)$ as a function of the external load $F$ at different overpressures. The aim of this study was to see how the diameter of the overpressurized straw behaved with variation in load. Due to the low thickness and the internal stress forces of the material, the straw could not maintain a circular shape without pressurization and should have a slightly ellipsoidal shape. Pressurization of the straw could return it to a strictly circular shape. 

The diameter of the straw began to increase as a function of pressure, as shown in Fig. \ref{fig:diam2}. Since a cross-section of the straw tube is circular and the internal and outer diameters are too close to each other, it is possible to consider just a circle to calculate its radial deformation. According to Hooke's law, the circle length relative increment (strain/elongation of straw cross-section) $\varepsilon_c$ could be calculated using equation \eqref{eq:strain}. The circle perimeter is proportional to its diameter, so the variation of diameter $\delta D$ for the thin-walled straw tube is proportional to the length of straw circle variation in the first approximation, and using equation \eqref{eq:stress_t}, it could be calculated as:
\begin{equation} \label{eq:dia_press}
    \delta D = \varepsilon_c D = \frac{\sigma_t}{E_m} D = \frac{D^2}{2T} \frac{p}{E_m}
\end{equation}

The variation of the straw tube diameter $D$ along its length was already shown in Fig. \ref{fig:longdiavar} for different overpressure. Figure \ref{fig:diaVSpress} presents the dependence of the straw tube diameter on the overpressure only in the middle of the straw. Note that uncertainties for the straw diameter variation shown in Fig. \ref{fig:diaVSpress2} are obtained from the distribution in Fig. \ref{fig:diam2}. These distributions in Fig. \ref{fig:diaVSpress} could be fitted with function \eqref{eq:dia_press}.

    \begin{figure}[!htbp]
    	\centering
    	\subfloat[]{\includegraphics[width=0.5\textwidth, keepaspectratio]
    		{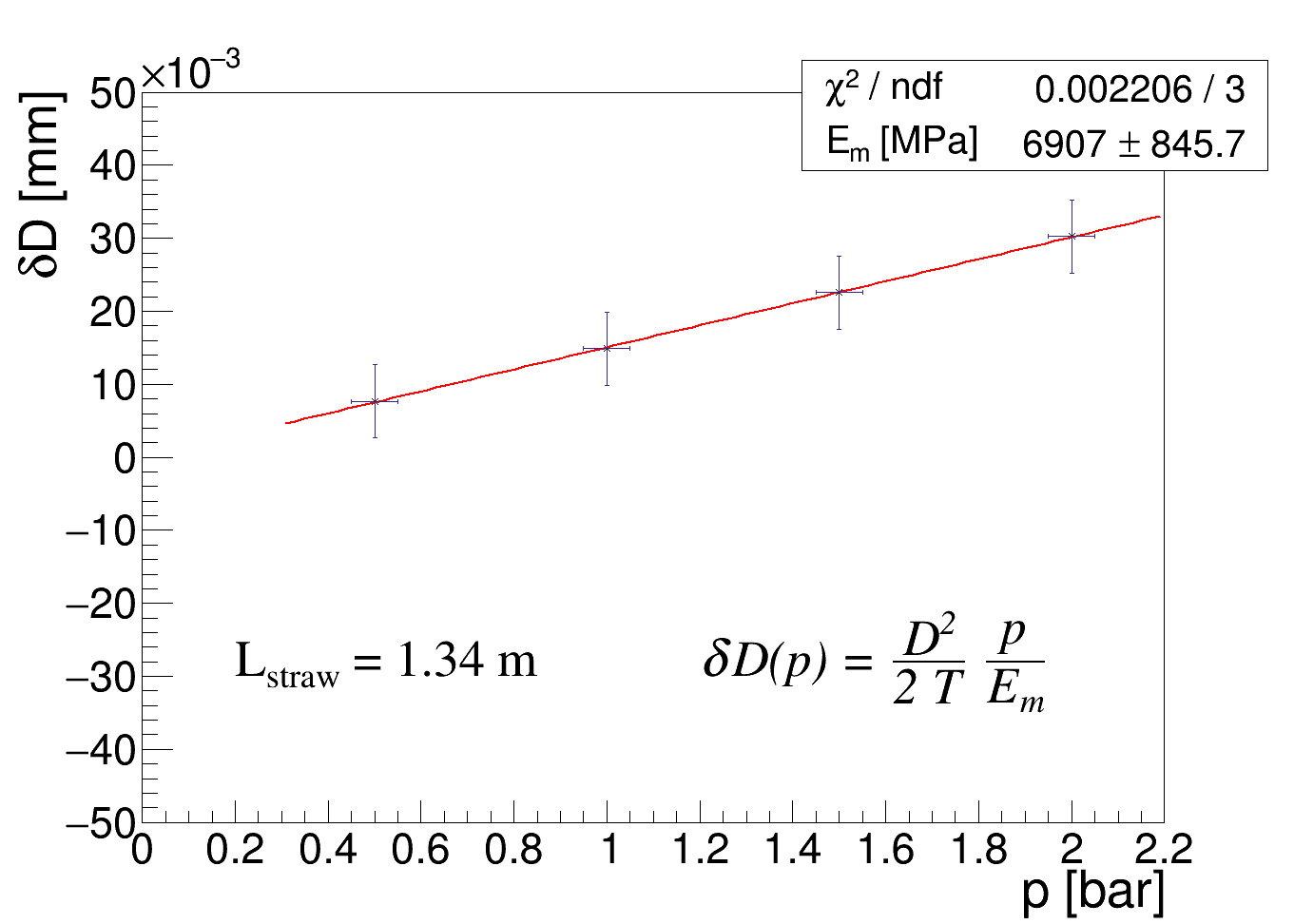} \label{fig:diaVSpress1}}
    	\subfloat[]{\includegraphics[width=0.5\textwidth, keepaspectratio]
		{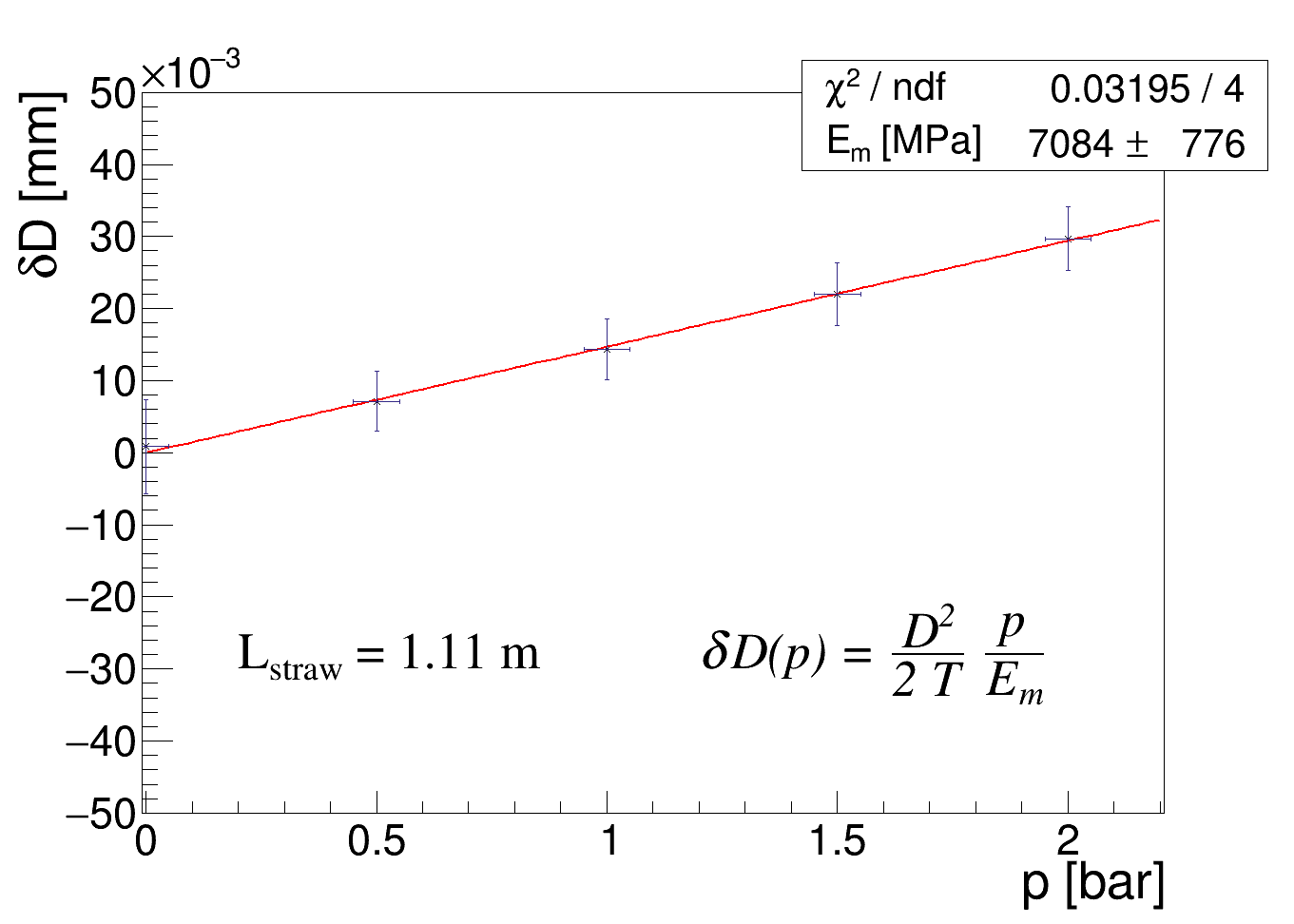} \label{fig:diaVSpress2}}
    	\caption{\small Straw diameter dependence of the over-pressure delivered to the straw: on the center of the 1.34-m-long straw (a) and the mean values of the diameter scanning along the straw (b) according to Figure {\ref{fig:diam2}}.}
    	\label{fig:diaVSpress}
    \end{figure}

According to these results, the diameter expansion caused by overpressure could be estimated at $16.1 \pm 2.0 ~\mu$m per 1 bar.

In contrast, an axial external load could decrease the diameter of the straw. Again, the thin-wall straw tube diameter variation $\delta D$ is proportional to the length of straw circle variation in the first approximation, and it could be calculated as:
\begin{equation} \label{eq:dia_load} 
    \delta D = \varepsilon_c D = \frac{\sigma_t}{E_m} D = \frac{\sigma_z}{-\nu}\frac{D}{E_m} = -\frac{F}{\nu S_{cr}}\frac{D}{E_m} \approx -\frac{mg}{\nu \pi TD}\frac{D}{E_m} = -\frac{g}{\pi T \nu}\frac{m}{E_m}
\end{equation}

The results of the diameter dependence on the axial external load for 0.5-, 1.0- and 1.34-m-long $\phi 5 mm \times t 12 \mu$m straws with the 0.5- and 1.0-bar overpressures applied are shown in Fig. \ref{fig:diaVSload}. These graphs were approximated by function \eqref{eq:dia_load}. 

    \begin{figure}[!htbp]
    	\centering
    	\subfloat[]{\includegraphics[width=0.33\textwidth, keepaspectratio]
    		{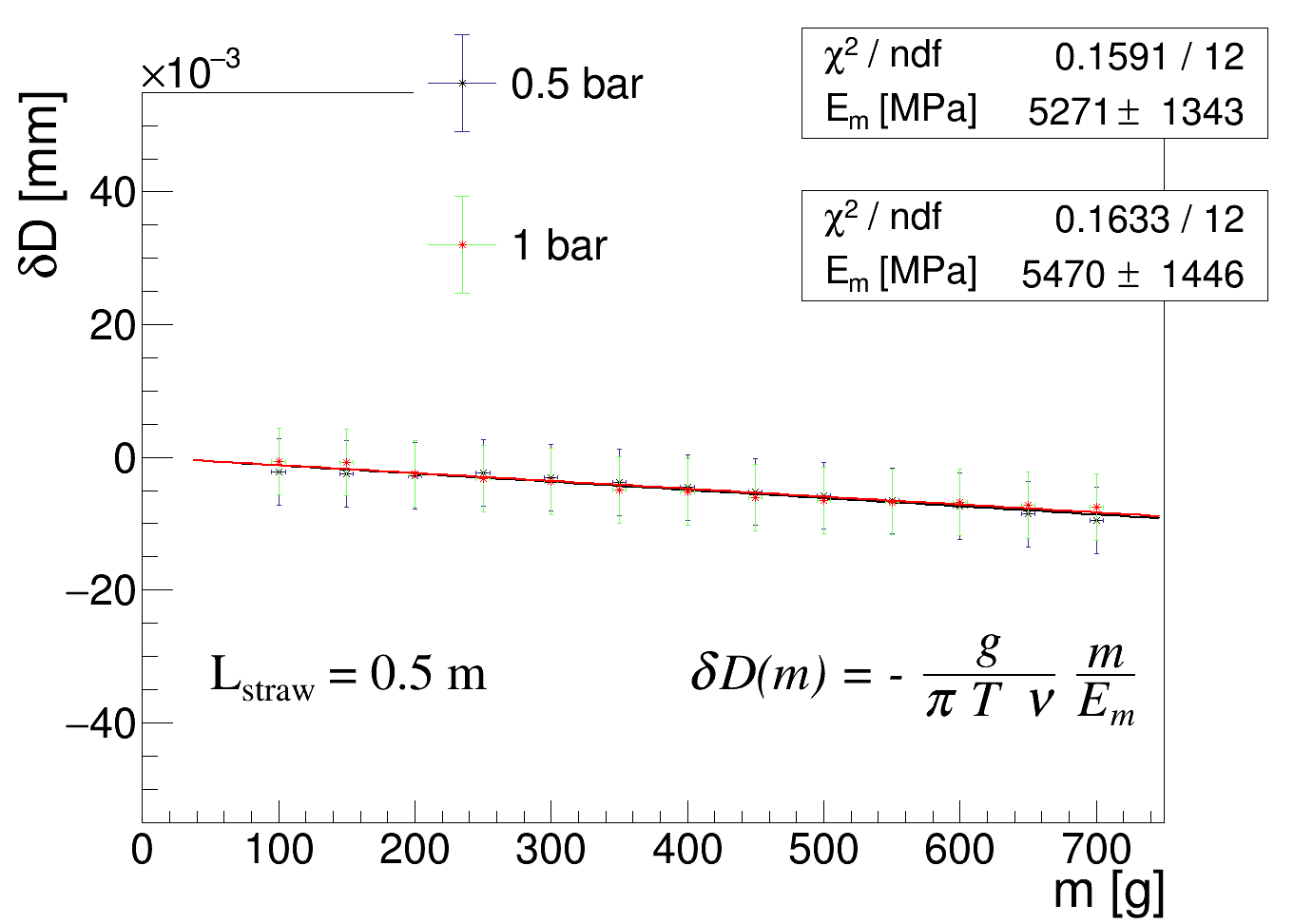} \label{fig:diaVSload1}}
    	\subfloat[]{\includegraphics[width=0.33\textwidth, keepaspectratio]
    		{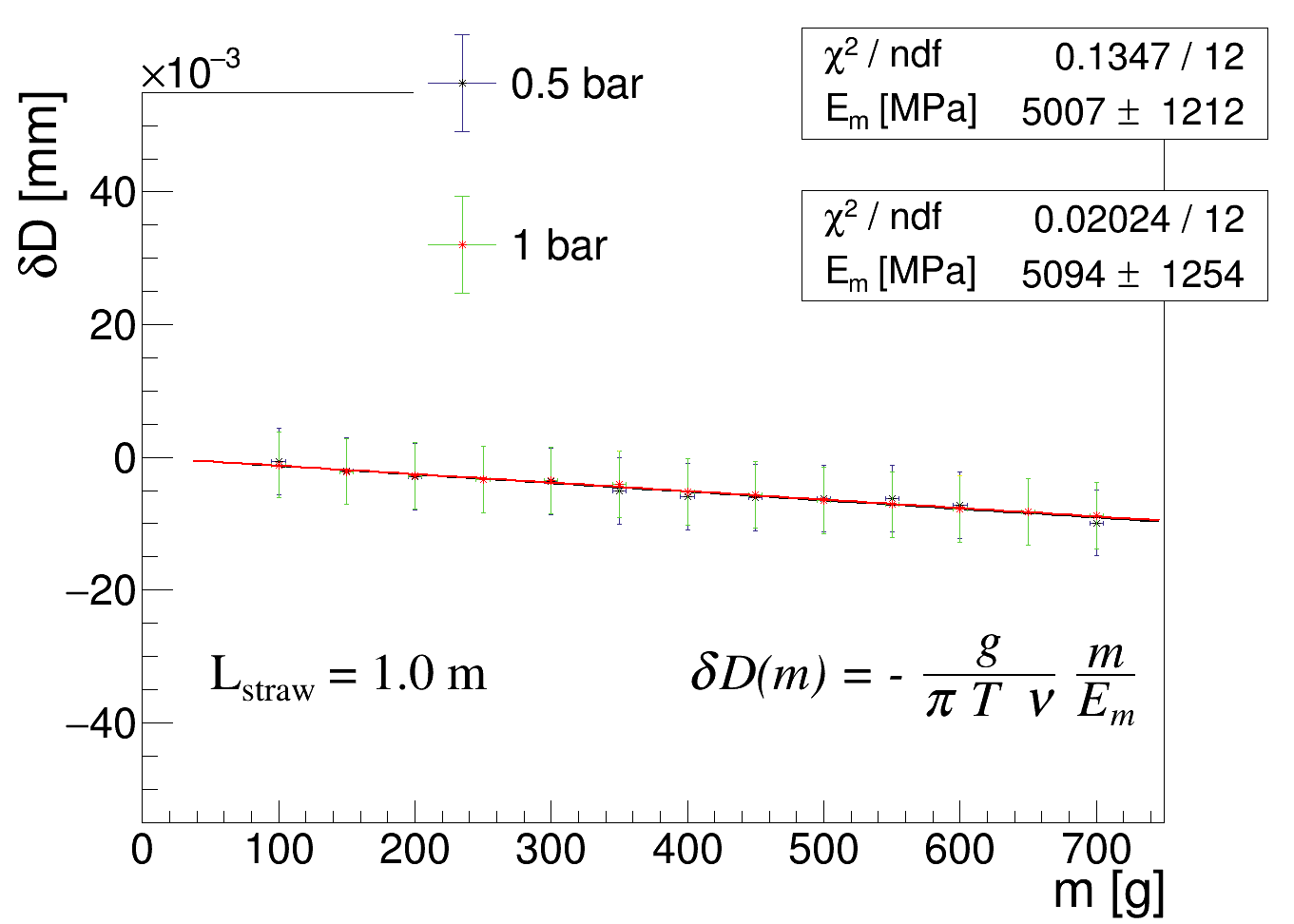} \label{fig:diaVSload2}}
    	\subfloat[]{\includegraphics[width=0.33\textwidth, keepaspectratio]
    		{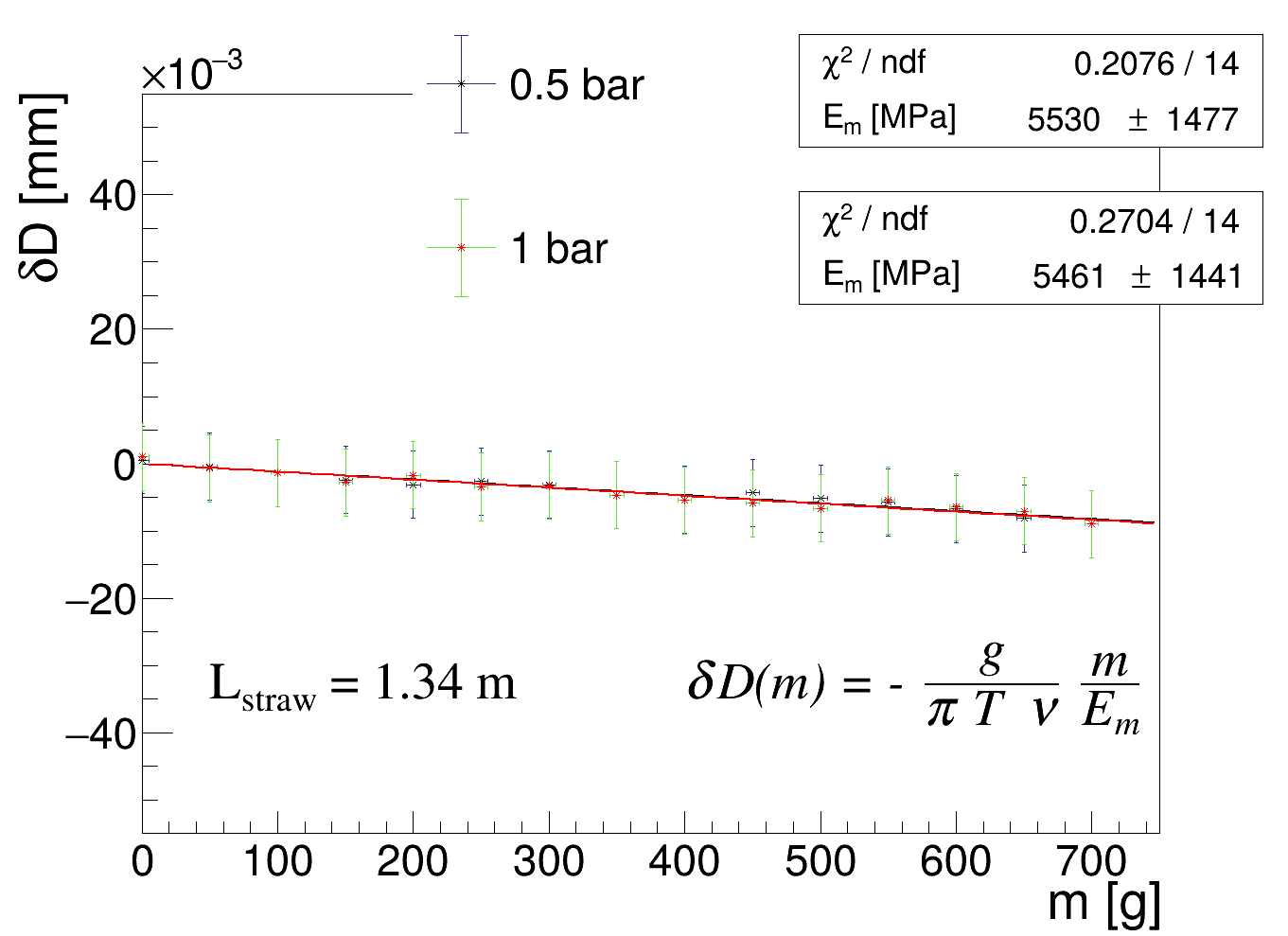} \label{fig:diaVSload3}}
    	\caption{\small Straw tube diameter dependence on the external load for different overpressure: 0.5 bar and 1.0 bar for the 0.5- (a), 1.0- (b) and 1.34-m-long (c) straw tubes}
    	\label{fig:diaVSload}
    \end{figure}

The graphs in Fig. \ref{fig:diaVSload} show that the straw diameter decreases in proportion to the external load in the whole scale up to 700 g. The decrement of the straw diameter is about $1.3 \pm 0.3$ $\mu$m per 100 g, and its behavior does not depend on the straw length or the overpressure up to 1 bar.

\subsection{Dependence of the straw tube sagging and bending on the external load}

Generally, straws lose their straightness due to bending. Bending was due to the following reasons: the uniqueness of production and the fact that straws could follow the tension map once pressurized. The action of gravitation on a straw tube should result in its sagging. A combination of bending and sagging could be observed in the horizontal orientation of the straw.

The curvature of the wire installed into the straw and the straw deformation under sagging and bending could differ, thus introducing displacement of the wire from the center of the straw as a function of the straw length. This fact could create additional uncertainties in particle track recognition. Usually, spacers could be used to solve this problem, but it also would be a reason to prevent the propagation of charge in this area. This shows the importance of studying straw bending and sagging. These measurements were done on setup \#2.

The bending study was carried out with the 1.34-m-long straw installed vertically with only the top end fixed. The results for these measurements are presented in Fig. \ref{fig:bending}. 

    \begin{figure}[!htbp]
    	\centering
    	\subfloat[]{\includegraphics[width=0.5\textwidth, keepaspectratio]
    		{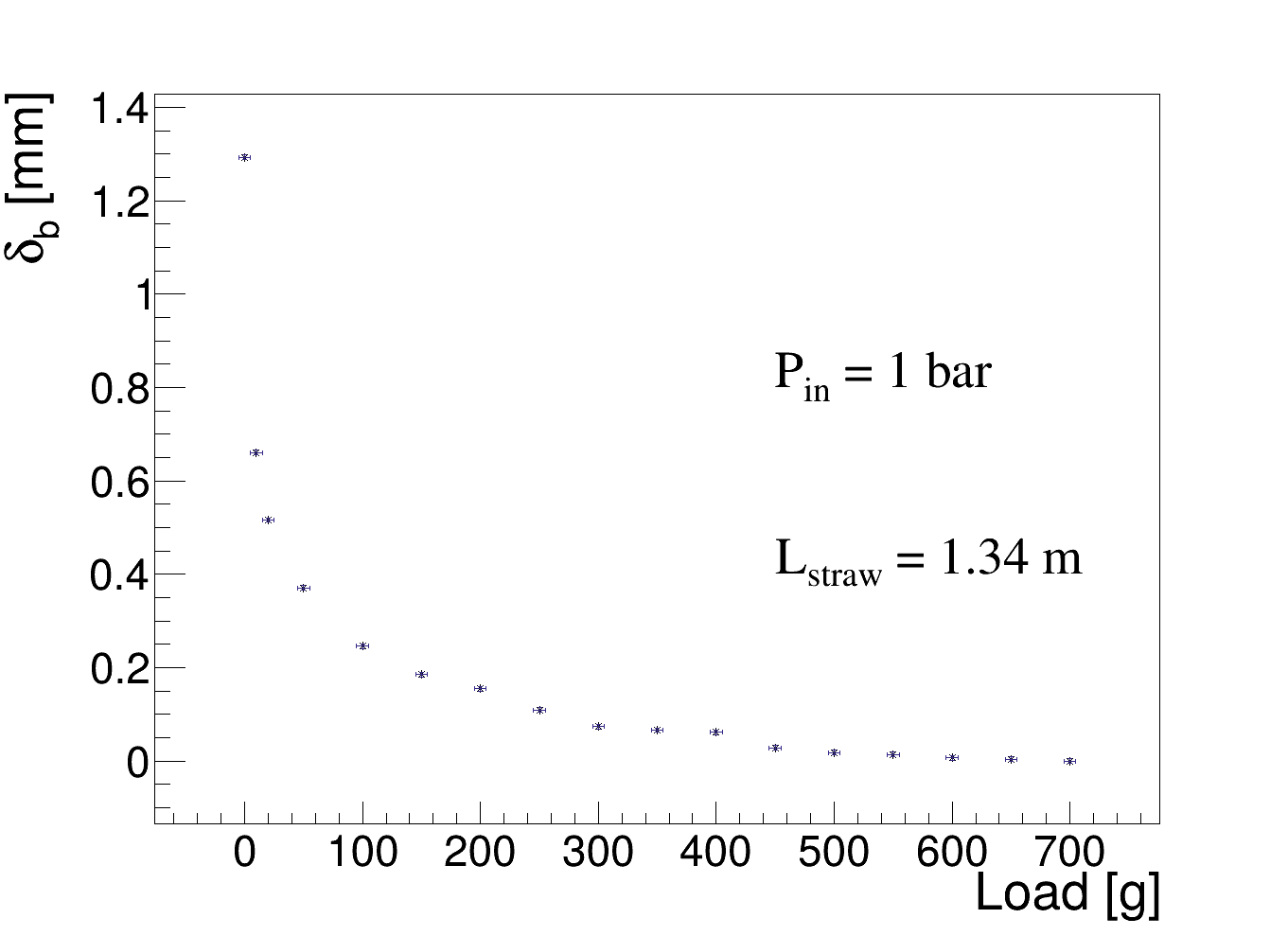} \label{fig:bending}}
    	\subfloat[]{\includegraphics[width=0.5\textwidth, keepaspectratio]
    		{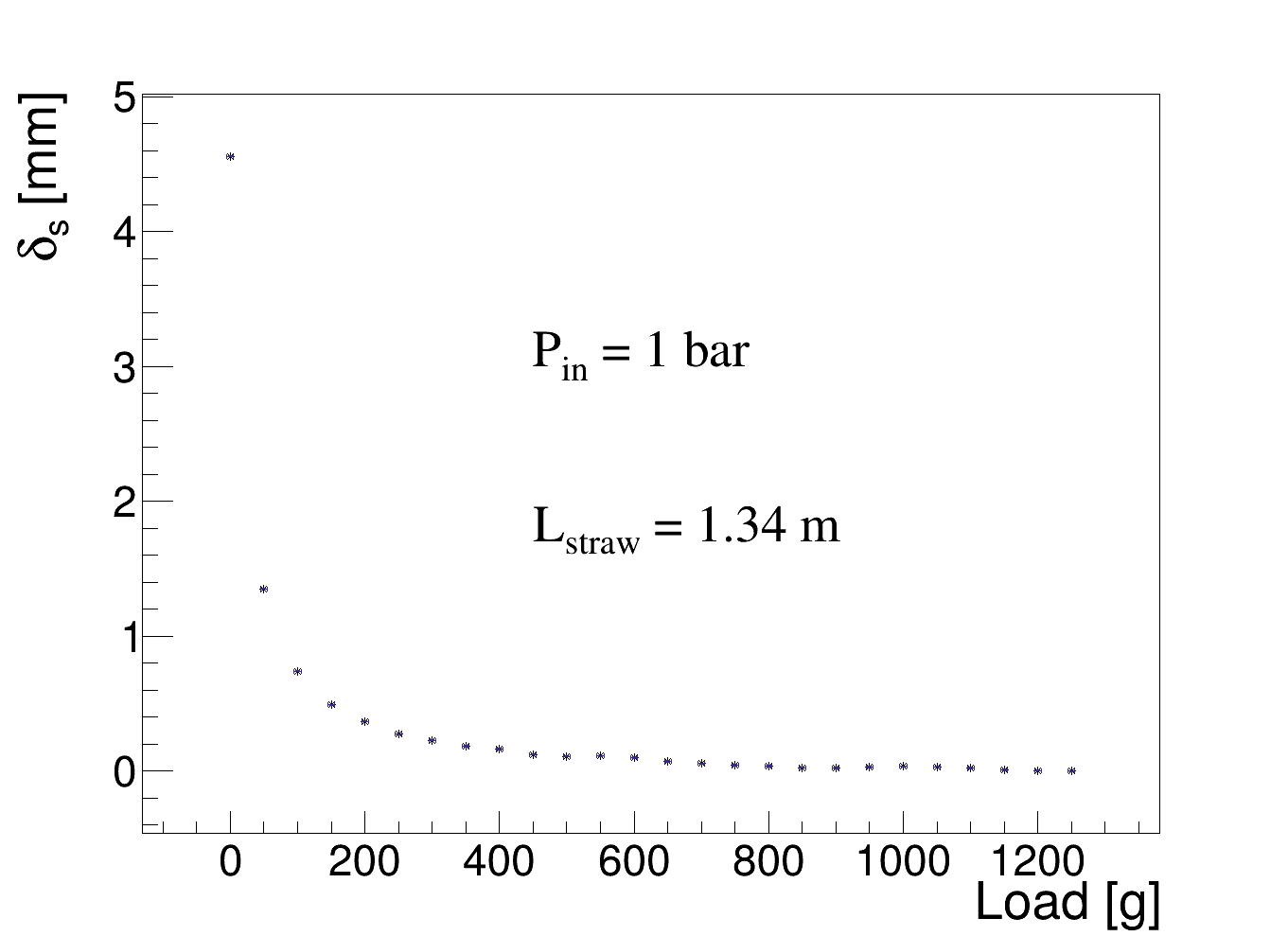} \label{fig:sagging}}
    	\caption{\small Displacements caused by the bending (a) and the sagging (b) as a function of the axial external load for a 1.34-m-long $\phi 5 mm \times t 12 \mu$m straw tube }
    	\label{fig:bendsang}
    \end{figure}

The diameter changed with the axial load as well and somehow affected measurements, but its values were small relative to the bending ones. For instance, the diameter changes by 0.04 mm at loads from 100 g to 400 g or would give an additional 0.02 mm for the edge shift observed on the laser micrometer. At same time, a relative displacement caused by bending for the same area was 0.1 mm.

The sagging of the 1.34-m-long straw was studied in the horizontal position of setup \#2, and, again, only one straw side was fixed while the other side was axially loaded. This study is illustrated in Fig. \ref{fig:sagging}. Note that the 1-bar overpressure was delivered into the straw to keep its shape during these measurements.

A load of 200 g is a minimal axial external load to the straw to hold the straightness of the 1.34-m-long $\phi 5 mm \times t 12 \mu$m straw at the level of 200 $\mu$m in both cases.

\section{Conclusions}
The station for the production of $\phi 5 mm \times t12 \mu$m straw tubes was successfully developed. About 100 straws were fabricated to show production stability. The straw length was 1.34 m as required by the COMET STS detector design.

Two different test setups were created, and different techniques were used for each setup to carry out the mechanical testing of the tubes. The straw tube properties (displacement and diameter changes under pressure and load, bending, and sagging) were studied in detail.

The axial displacement study showed that the straws could be loaded with up to 700 g and remain within the stretching linear area. The Young module was calculated to be about $6.57\pm 0.89$ GPa.

The diameter of the straw depended on the overpressure. It was found that the diameter of the straw expands by $16 \mu$m per 1 bar.

The straw elongation caused by internal over-pressure proportionally depends on the internal over-pressure. As expected, radial stiffness for a 0.5-m-long straw was twice higher than for a 1-m-long straw.  

In contrast to the overpressure, the diameter of the straw decreases under the axial external load on the straw, and the diameter decrement behavior does not depend on the straw length or the overpressure of up to 1 bar applied to the straw. The straw diameter shrink rate was found to be 1.3 $\mu$m per 100 g.

Both bending and sagging studies showed that a load of at least 200 g is needed to keep the straightness of the 1.34-m-long straw better than $200 \mu$m.

It was experimentally confirmed that straws with this design could withstand the 3-bar overpressure. The preliminary estimation of the gas leakage is $1.56 \pm 10^{-4}$ ccm per 1 m for the $\phi 5 mm \times t 12 \mu$m straw tube where the 2-bar overpressure was applied with pure Ar gas. These two facts could allow the straws with a wall of 12 $\mu$m to be used for COMET Phase-II with the 1-bar pressure to operate in a vacuum.

\section{Acknowledgments}\label{thanks}
This work was supported by the Shota Rustaveli National Science Foundation of Georgia (SRNSFG), grant no. [PHDF-19-3553], by the EU Horizon 2020 Research and Innovation Programme under the EXCELLENT SCIENCE - MSCA, PROBES, GA 101003460 and by JSPS KAKENHI 21H04482, 16KK0108, and 16H02192.

The authors are grateful to GTU (Tbilisi) and JINR (Dubna) for supporting this study.

We also thank Dr. Hans Danielsson (Technical Coordinator \& Deputy EXSO of NA62) for the straw seam study by using an imaging microscope.

We are grateful to Guram Tushmalishvili for useful discussion in the field of Material Engineering, and to Iliya Chokheli for supporting this research by providing the design of CAD models and developing them for the test setup.

\bibliography{
    bib_cometstsmech5mm.bib
}
\end{document}